\documentclass[twocolumn]{aastex7}
\usepackage{bm}
\usepackage{ae,aecompl,amsmath}

\begin{document}

\title{On the orbital evolution of binaries with polar circumbinary disks}
 
\author[orcid=0000-0002-4489-3491,sname='Chen']{Cheng Chen}
\email[show]{cchen@ubishops.ca}
\affiliation{School of Physics and Astronomy, University of Leeds, Sir William Henry Bragg Building, Woodhouse Ln., Leeds LS2 9JT, UK}
\affiliation{Department of Physics and Astronomy, Bishop's University, 2600 Rue College, Sherbrooke, QC J1M 1Z7, Canada}

\author[orcid=0000-0001-8974-0758,sname='Armitage']{Philip J. Armitage}
\email[hide]{philip.armitage@stonybrook.edu}
\affiliation{Center for Computational Astrophysics, Flatiron Institute, New York, NY 10010, USA}
\affiliation{Departmant of Physics and Astronomy, Stony Brook University, Stony Brook, NY 11794, USA}

\author[orcid=0000-0002-2137-4146,sname='Nixon']{C. J. Nixon}
\email[hide]{C.J.Nixon@leeds.ac.uk}
\affiliation{School of Physics and Astronomy, University of Leeds, Sir William Henry Bragg Building, Woodhouse Ln., Leeds LS2 9JT, UK}

\begin{abstract}
Binaries occur in many astrophysical systems, from young protostellar binaries in star forming regions to supermassive black hole binaries in galaxy centers. In many cases, a circumbinary disk of gas forms around the binary with an orbit that may be misaligned to the binary plane. Misaligned disks around nearly circular binaries evolve into disks that are either aligned or counteraligned with the binary orbit. However, if the binary is sufficiently eccentric, then it can be more likely that the disk ends up in a polar-aligned configuration in which the disk angular momentum vector aligns with the binary eccentricity vector. We use Smoothed Particle Hydrodynamics simulations, evolved to an approximate steady state under mass injection, to determine the orbital evolution of a binary with a polar-aligned disk for a range of binary-disk parameters. We find that, in all of the cases we have simulated, the binary shrinks with time. The decay rate is larger than for binaries surrounded by aligned or retrograde disks with matched disk parameters. The rate of shrinkage is largely unaltered by the size of the sink radii employed for the binary stars, but for small enough sink radii some of the models exhibit long-lived polar circumprimary disks, which are continually fed mass from the circumbinary disk. We discuss our results in the contexts of planet formation in young polar-aligned disks and merging supermassive black holes in galaxy centers.
\end{abstract}


\section{Introduction}
Binary stars are common in the universe. About half of all solar like stars are in binaries or higher multiple systems \citep[e.g.][]{Raghavan2010} and an even higher fraction of massive stars are in binaries \cite[e.g.][]{Sana2012}. Binary systems of stellar mass compact objects may also form out of dynamical capture processes in AGN disks  \citep[e.g.][]{Wang:2024}, while supermassive black hole binaries form in galaxy centers following the merger of two galaxies \citep{Begelmanetal1980}.

Binary stars may be accompanied by a circumbinary disk (CBD) of gas. CBDs differ from standard disks due to the gravitational torques acting between the binary and the disk orbits, typically generated through Lindblad resonances. For disks  that are aligned to the binary orbital plane, these torques serve to transfer energy and angular momentum from the binary to the disk material \citep{Lin:1986,Pringle1991,Artymowicz1991}. The balance between the angular momentum flux associated with these torques and the flux of angular momentum from accretion of material on to the binary determines how the binary orbit evolves \citep[][and references therein]{Heath2020}. For disks  with strong viscous torques and weak resonant torques, the excess angular momentum transferred to the binary through accretion can lead to expansion of the binary with time \citep[as shown by the simulations presented in, e.g.,][]{Miranda2017,Tang2017,Munoz2019,Moody2019}. For thin disks  in which the viscosity is not anomalously strong, the expectation is that the binary contracts with time \citep{Tiede2020,Heath2020} as originally envisaged by, e.g., \cite{Pringle1991,Artymowicz1994}. 

For misaligned CBDs the disk orbits also undergo precession induced by the gravitational potential of the binary \citep{Larwood:1997}. The long-term evolution of the disk is then to move towards minimum energy states that remove the precession torque. For circular binaries these states correspond to aligned and counter-aligned orbits \citep{Nixonetal2011b}, and in the counteraligned case the Lindblad resonances are absent leading to unrestricted accretion on to the binary \citep{Nixonetal2011a}. Thus counteraligned disks  lead to shrinkage of the binary orbit, typically by increasing the eccentricity of the binary while also reducing the semi-major axis. 

For eccentric binaries, the aligned and counter-aligned states are supplemented by a third stable state; a polar-aligned state in which the disk aligns to the binary eccentricity vector \citep{Aly2015,Martin:2017,Zanazzi:2018}. The inclusion of eccentricity introduces several new complexities. For aligned CBDs, the torque that the binary exerts on the disk depends on the binary’s eccentricity and the accretion torque depends on eccentricity. This dependence arises directly because eccentricity changes the binary’s specific angular momentum.
It also arises indirectly through inner-disk dynamics: the cavity is often eccentric, and accretion proceeds in time‑dependent streams whose phase affects the torque. As such, the response of an eccentric binary to a CBD is complex and highly parameter dependent \citep[e.g.][]{D'Orazio:2021}. The retrograde case is also more complex due to the return of orbital resonances between the binary and the disk; decomposing the eccentric binary potential into rigidly rotating bars leads to some components of the potential which rotate opposite to the binary orbit, and thus prograde with respect to the retrograde CBD \citep{Nixon2015}. These resonances are typically weaker than their prograde counterparts, so accretion is not significantly slowed but does become strongly oscillatory.

In this paper, we focus on the polar-aligned state and present a set of simulations aimed at determining the response of the binary to polar-aligned disks. We aim to determine if the binary orbit shrinks or expands in the polar-aligned case. We therefore present simulations with different disk-binary parameters. We first describe the setup of the simulations and the parameter domain we explore in Section~\ref{sta}. We describe our results in Section~\ref{sim1}, and specifically we consider smaller accretion radii of the binary in Subsection~\ref{sim2} and study the effect of the binary mass ratio in Subsection~\ref{sim3}. Finally, we present our discussion and conclusion in Sections~\ref{diss} and Section~\ref{con}.

\section{simulation setup of binaries with polar circumbinary disks}
\label{sta}
In this section, we outline the code and simulation setup that we employ, along with the parameter space that we explore. We employ the 3D SPH code \textsc{phantom} \citep{Price2018}. This code has been widely used for exploring circumbinary systems \citep[starting with][]{Nixon2012}. Our simulation considers a binary system with a total mass of $M_1+M_2 = M_{\rm b}$ in a eccentric orbit with eccentricity $e_{\rm b}$, and an initial semi-major axis $a_{\rm b}$. We present simulations with $e_{\rm b} = 0.4$, $0.6$ and $0.8$ with the binary mass ratio $q_{\rm b} = M_2/M_1 = 1$. For $e_{\rm b} = 0.4$, we also present results for a case with the binary mass ratio $q_{\rm b}=1/3$. The accompanying polar CBD is placed in a circular orbit but initially perpendicular to the binary orbital plane with a mass $M_{\rm d}=10^{-6} M_{\rm b}$. We adopt such a small disk mass so that the stationary inclination angle for the disk is close to the polar, $90^\circ$ inclination to the binary orbit.\footnote{For small disk masses we expect that our results can be suitably rescaled by the disk mass, but for large enough disk masses this will not be the case due to several effects, including the change in the stationary inclination and the effects of self-gravity in the disk dynamics.}

The initial disk extends from $R_{\rm in} = 3a_{\rm b}$ to $R_{\rm out} = 10a_{\rm b}$. We take the disk angular semi-thickness $H/R = 0.1$ at $R = R_{\rm in}$, and we take the sound-speed of the gas to be constant with radius, which results in $H/R \propto R^{1/2}$. We perform simulations with an initial particle number of $N_{\rm p,ini}$ =  10$^{6}$.

We impose an outer boundary for the simulation at $R_{\rm out}$, which means that any particle with $R > R_{\rm out}$ is removed from the simulation. The accretion radii of both stars, $R_{\rm acc}$, are initially set at 0.4$a_{\rm b}$ as this facilitates a faster development of a steady state; we subsequently decrease the accretion radii to explore the flow in and around the binary.

We take the initial surface density $\Sigma$ to be given by \citep[][see also the Appendix of \citealt{Drewes2021}]{Nixon2021}
\begin{equation}   
\resizebox{0.47\textwidth}{!}{$
\Sigma (R)=\left\{
\begin{aligned}
& \frac{\dot{M}_{\rm add}}{3\pi \nu (R)} \left[ 1 - \left(\frac{R_{\rm in}}{R} \right)^{1/2}\right]\frac{R_{\rm out}^{1/2}-R_{\rm add}^{1/2}}{R_{\rm out}^{1/2}-R_{\rm in}^{1/2}} &\ {\rm for} \ R\ \leq \ R_{\rm add} \\
& \frac{\dot{M}_{\rm add}}{3\pi \nu (R)}  \left[ \left(\frac{R_{\rm out}}{R} \right)^{1/2} -1 \right]\frac{R_{\rm add}^{1/2}-R_{\rm in}^{1/2}}{R_{\rm out}^{1/2}-R_{\rm in}^{1/2}} &\ {\rm for} \ R\ > \ R_{\rm add}
\end{aligned}
\right.
$}
\end{equation}
and we add mass to the disk at the rate $\dot{M}_{\rm add}$ at a radius $R_{\rm add} = 7a_{\rm b}$ so that the disk can reach a steady state around the binary. We hold the initial binary orbit to be fixed until this steady state is reached. After this phase, we allow the binary orbit to evolve to measure the response of the binary to the disk directly from the simulation.

The physical viscosity is modeled by a Navier-Stokes viscosity corresponding to a Shakura-Sunyaev disk viscosity parameter $\alpha_{\rm SS}=0.1$ \citep{SS1973}. The artificial viscosity applied in SPH simulations is typically composed of linear and quadratic terms. We consider a linear term ($\alpha_{\rm SPH}$) for each particle that is set using a modified version of the time-dependent Cullen-Dehnen switch \citep[][see \citealt{Price2018} for details]{Cullen2010} with a minimum value of 0.01 and a maximum value of unity, and a quadratic term where the value of $\beta_{\rm SPH}=2\alpha_{\rm SPH}$. This enables the numerical viscosity to operate where needed, while minimizing its influence in regions where it is not required \citep[see][for discussion]{Chen2025a}. 

\begin{table}
\centering
\caption{This table summarizes the simulation parameters. The first column lists the name of each model. The second and third columns indicate the mass ratio of the binary, and the binary eccentricity, respectively. The final two columns present the results of the binary's orbital evolution (the values represent ${\dot a}$ in units of $a_{\rm b}/T_{\rm b}$) with two different $R_{\rm acc}$ values and the binary's orbital evolution rate in the unit of $a_{\rm b}/T_{\rm b}$. For model D, their $R_{\rm acc}$ is equal to 0.1 times the effective Roche lobe of each star.}
\begin{tabular}{cccccc} 
\hline
\textbf{Model} & $q_{\rm b}\ (M_2/M_1)$   & $e_{\rm b
}$ & $R_{\rm acc}:
0.4 a_{\rm b}$   & $R_{\rm acc}:0.05 a_{\rm b}$\\
\hline
\hline
A  & 1 & 0.4  & -7.7$\times 10^{-9}$ &  -1.2$\times 10^{-8}$ \\
B  & 1 & 0.6   & -1.2$\times 10^{-8}$ &  -1.5$\times 10^{-8}$ \\
C  & 1 & 0.8  & -1.5$\times 10^{-8}$ &  -1.9$\times 10^{-8}$  \\
\hline
\textbf{Model} & $q_{\rm b}\ (M_2/M_1)$   & $e_{\rm b
}$ & $R_{\rm acc}:
0.4 a_{\rm b}$   & $R_{\rm acc}:0.1 R_{\rm RL}$\\
\hline
D  & 1/3 & 0.4   & -8.9$\times 10^{-9}$ & -8.1$\times 10^{-9}$  \\
\hline
\label{table1}
\end{tabular}
\end{table}

\section{Simulation results}
This section presents the outcomes of our SPH simulations. We collect our discussion of the results into three subsections corresponding to two different sizes for the accretion radii of the binary for an equal mass ratio (with different eccentricities presented together), and then we discuss a case with an unequal mass ratio. Specifically these are: (i) an equal mass binary with large accretion radii of $R_{\rm acc}=0.4 a_{\rm b}$ (Section~\ref{sim1}), (ii) an equal mass binary with small accretion radii of $R_{\rm acc}=0.025 a_{\rm b}$ (Section~\ref{sim2}), and (iii) an unequal mass binary with both large and small accretion radii, where the large radii are again $R_{\rm acc}=0.4 a_{\rm b}$ for both stars, and as the binary masses are now different we set the small radii to be fractions of their respective Roche lobe, i.e. $R_{\rm acc} = 0.1R_{\rm RL}$ (where the Roche lobe size is calculated assuming a circular binary\footnote{This is therefore equal to $R_{\rm acc} = 0.167R_{\rm RL,p}$ where $R_{\rm RL,p}$ is the Roche lobe size at the pericenter of the binary orbit as this case has $e_{\rm b} = 0.4$}).

\subsection{Disks with $R_{\rm acc}$ = 0.4 $a_{\rm b}$}
\label{sim1}
We first show disk models A -- C which have $H/R = 0.1$, $\alpha = 0.1$ and $q_{\rm b} = 1$ with $e_{\rm b} = 0.4$, 0.6 and 0.8, respectively. The accretion radii of the binary stars ($R_{\rm acc}$) are set to $0.4 a_{\rm b}$ and we let the disk evolve until it reaches a steady state. By the time the models reach steady states, the number of particles in the disk, $N_{\rm p}$, has increased to about 2$\times 10^6$ (model A), 2.6$\times 10^6$ (model B) and 3.2$\times 10^6$ (model C). As a result, $m_{\rm d}$ of each model also increases with increasing $N_{\rm p}$ at the steady state.

\begin{figure*}
    \centering
        \includegraphics[width=7.cm]{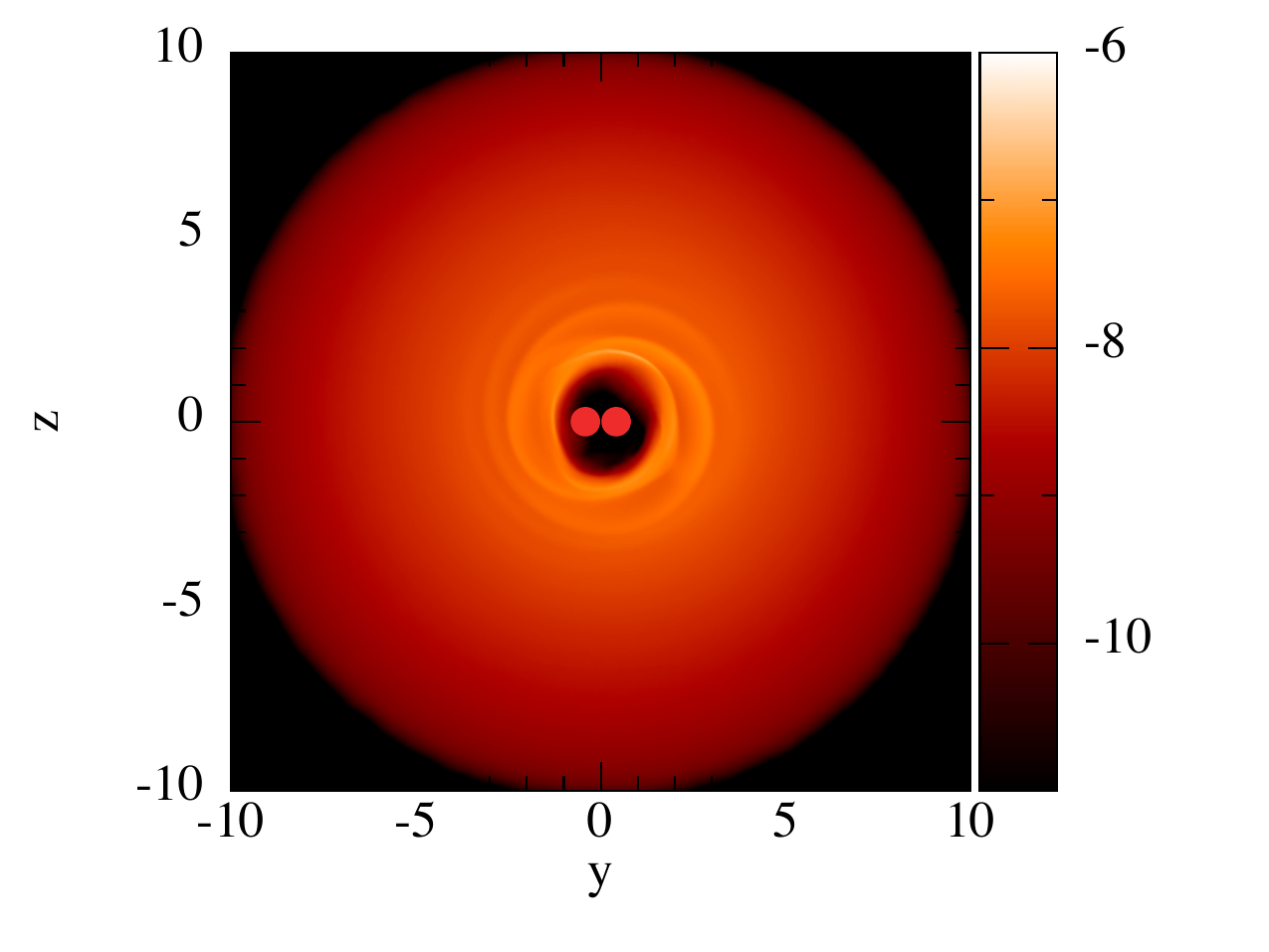}
        \includegraphics[width=7.cm]{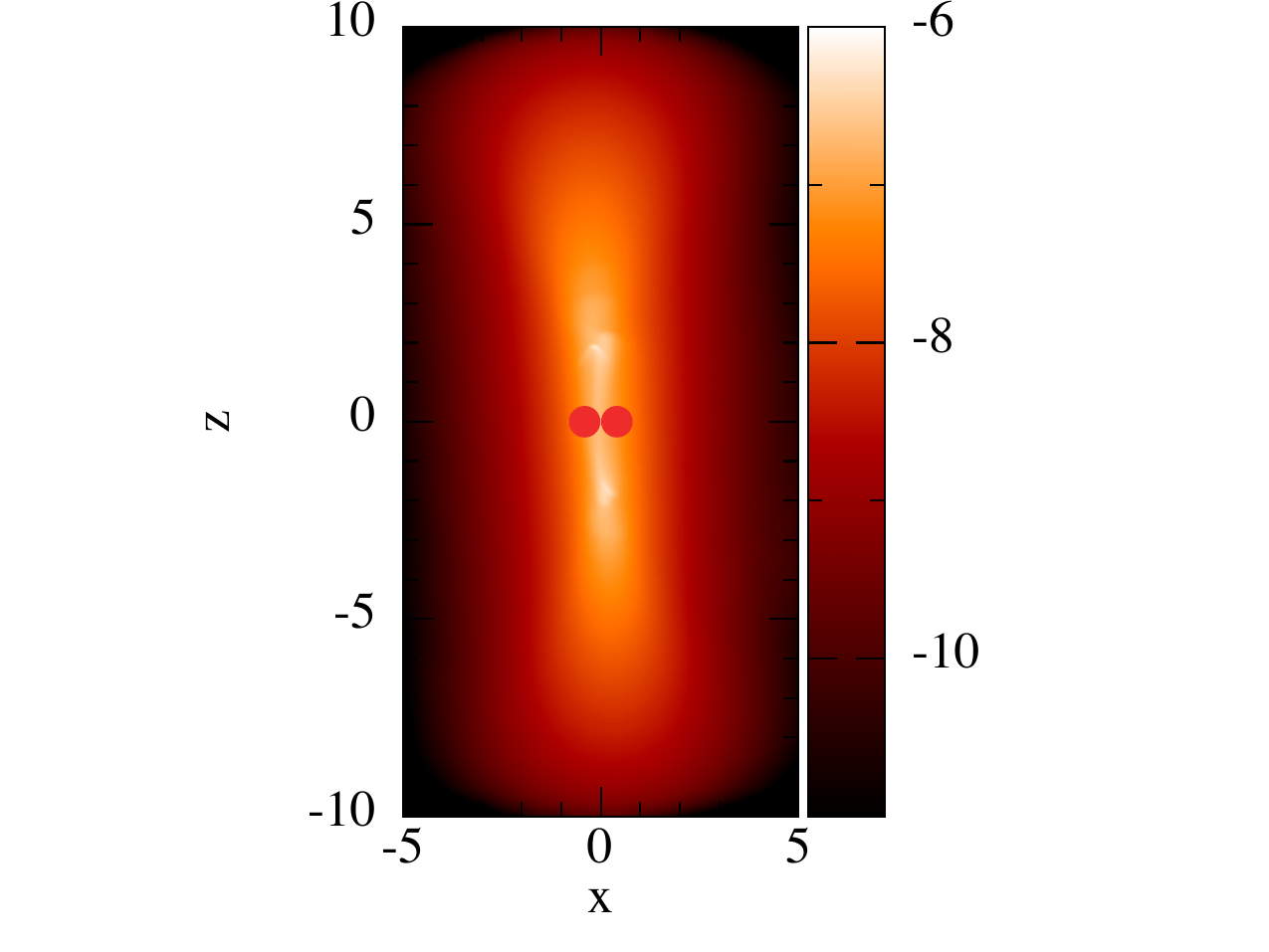}      
        \includegraphics[width=7.cm]{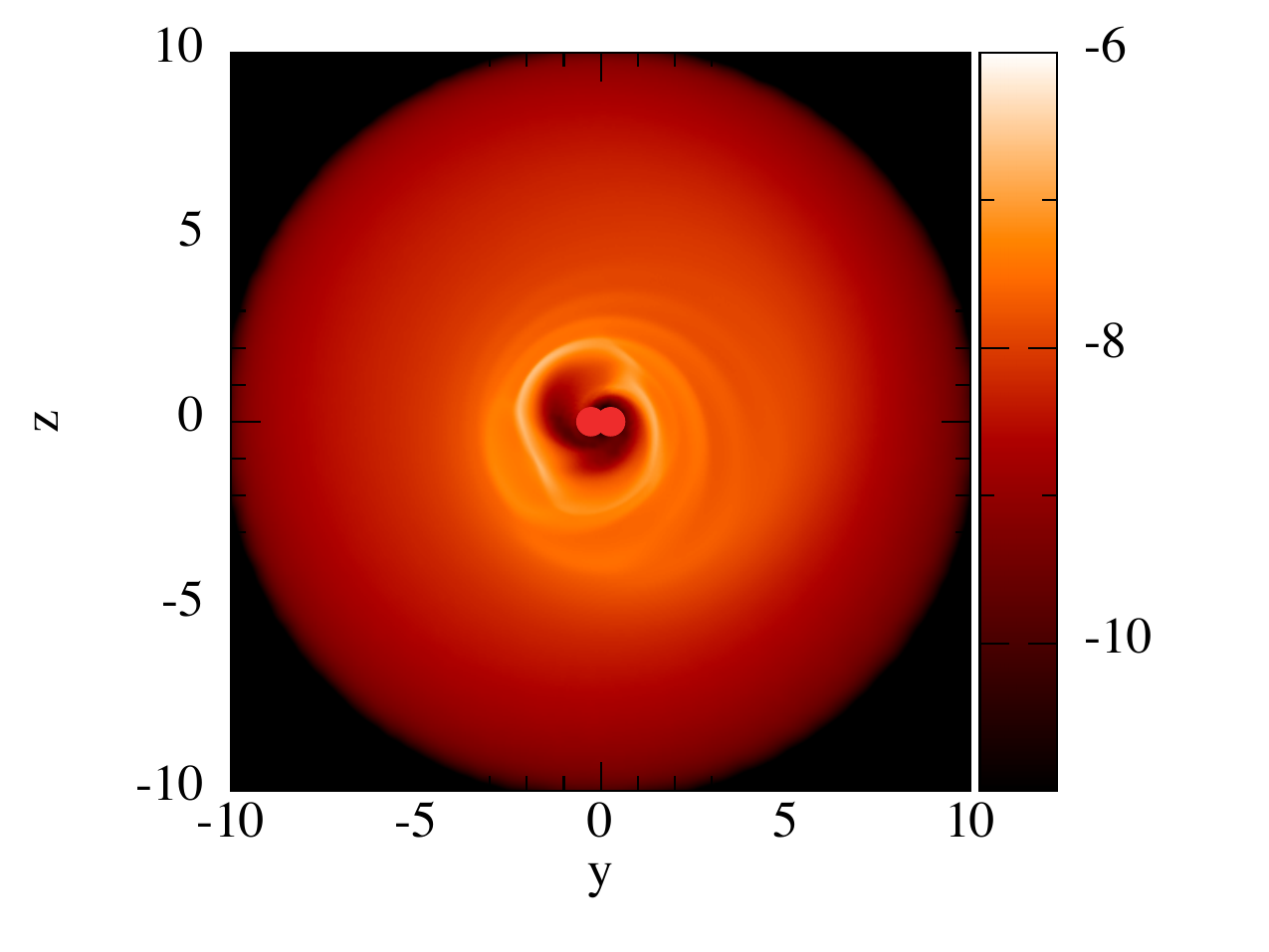}
        \includegraphics[width=7.cm]{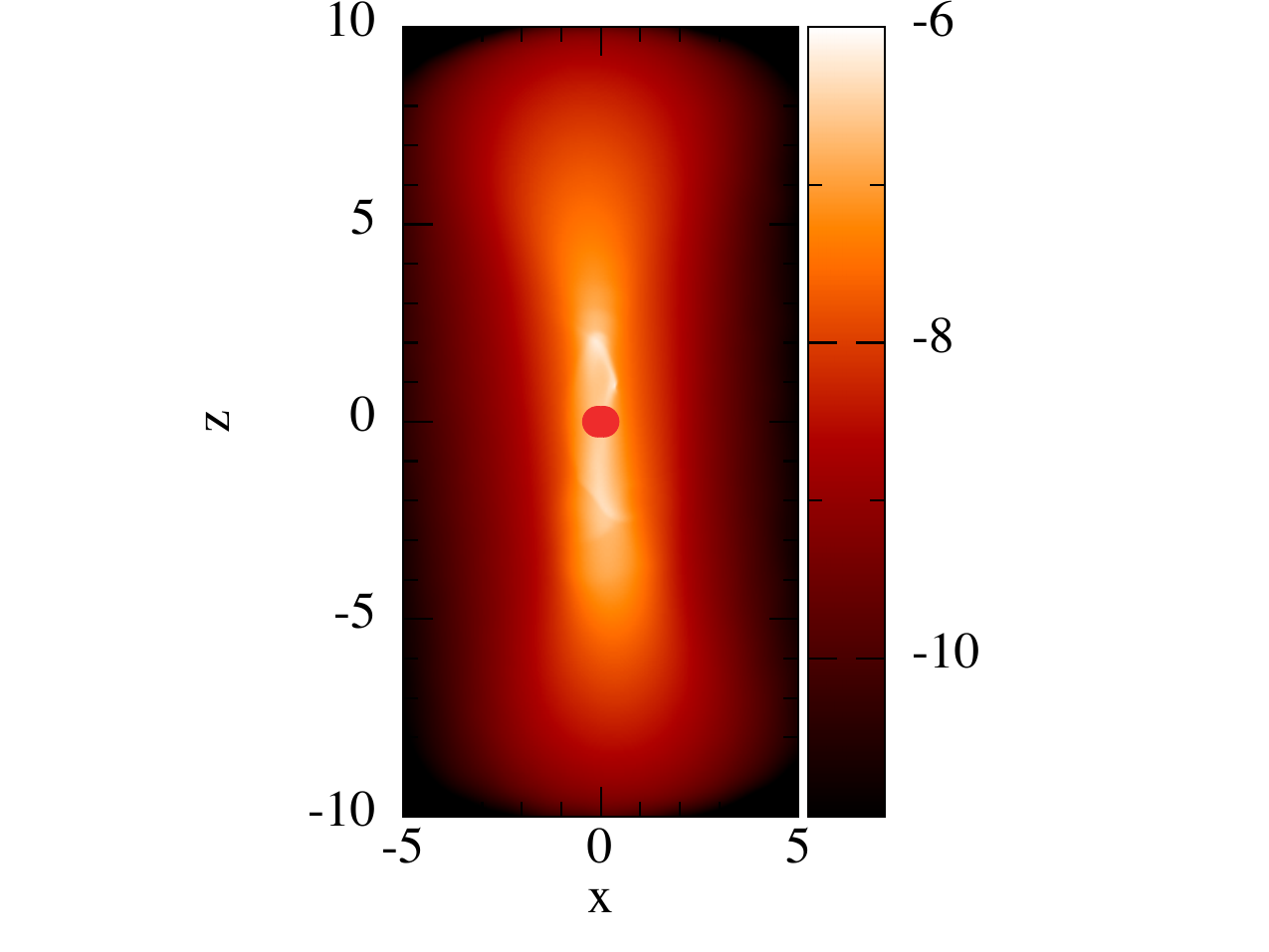}       
        \includegraphics[width=7.cm]{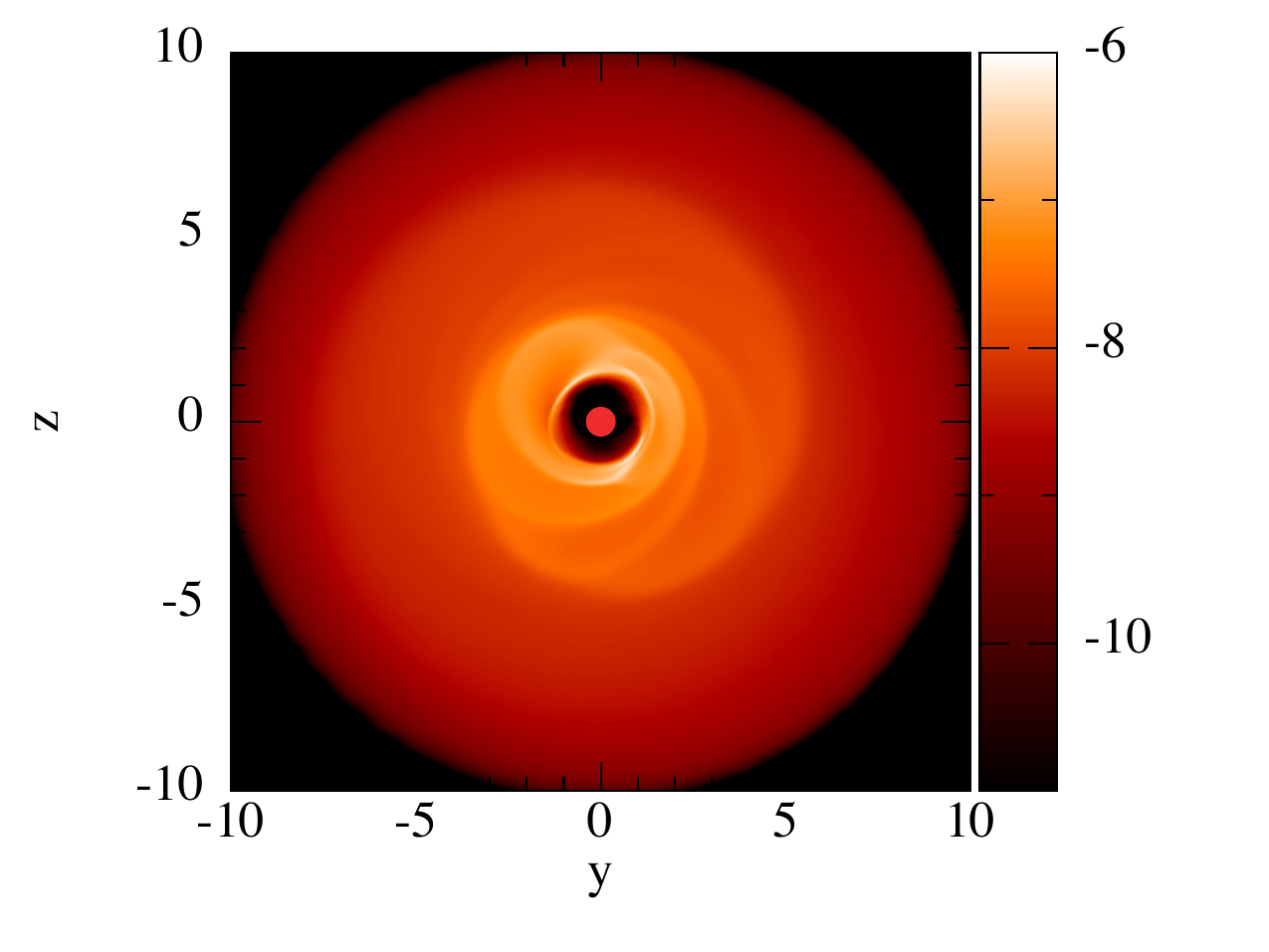}
        \includegraphics[width=7.cm]{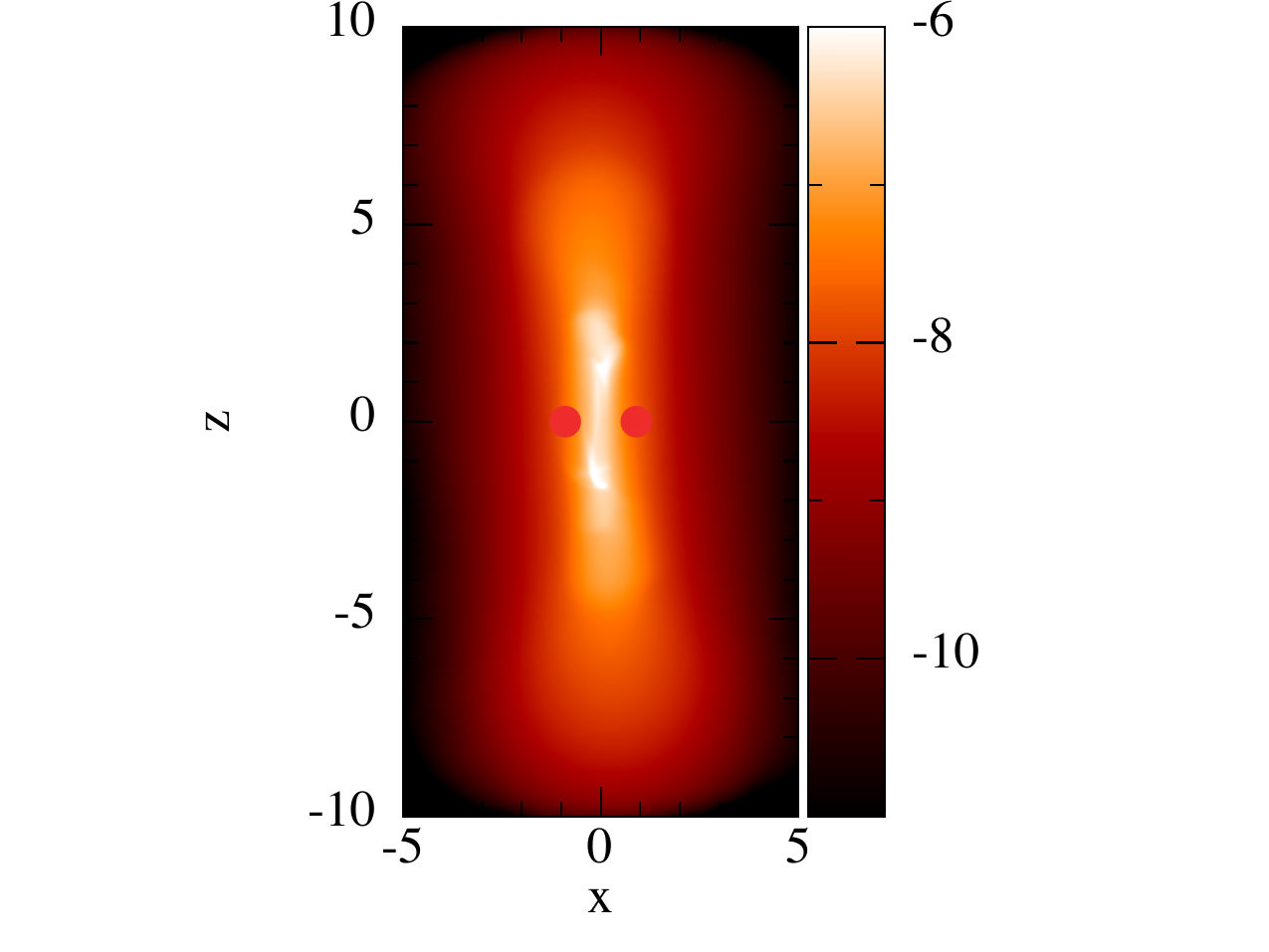}      
    \caption{Column density plots for the models A (top panels), B (middle panels) and C (lower panels) in the $y$-$z$ plane and the $x$-$z$ plane. The disks are shown once they have reached a steady state. Red circles mark the binary locations, with sizes equal to the accretion radii for each star. The unit of the axis is $a_{\rm b}$, and the color bars are the same for all of the panels with the density in arbitrary units.}
    \label{fig:1}
\end{figure*}

Fig~\ref{fig:1} shows the disks  in models A -- C once the steady state is reached with the left hand column showing the disk face on ($y$-$z$ plane) and the right column showing the disk edge on ($x$-$z$ plane); the binary orbits in the $x$-$y$ plane. The top two panels correspond to model A, the middle to model B, and the bottom to model C. The two red circles indicate the locations of the binary stars and the size of the circles is the size of the accretion radii (in some plots their location overlaps in projection). In left panels, we can see density waves are excited by resonances between the binary and the disk. In right panels, we can see that the three disks  are, as expected, aligned to nearly 90$^{\circ}$ in the steady state.

\begin{figure*}
    \centering
        \includegraphics[width=8.3cm]{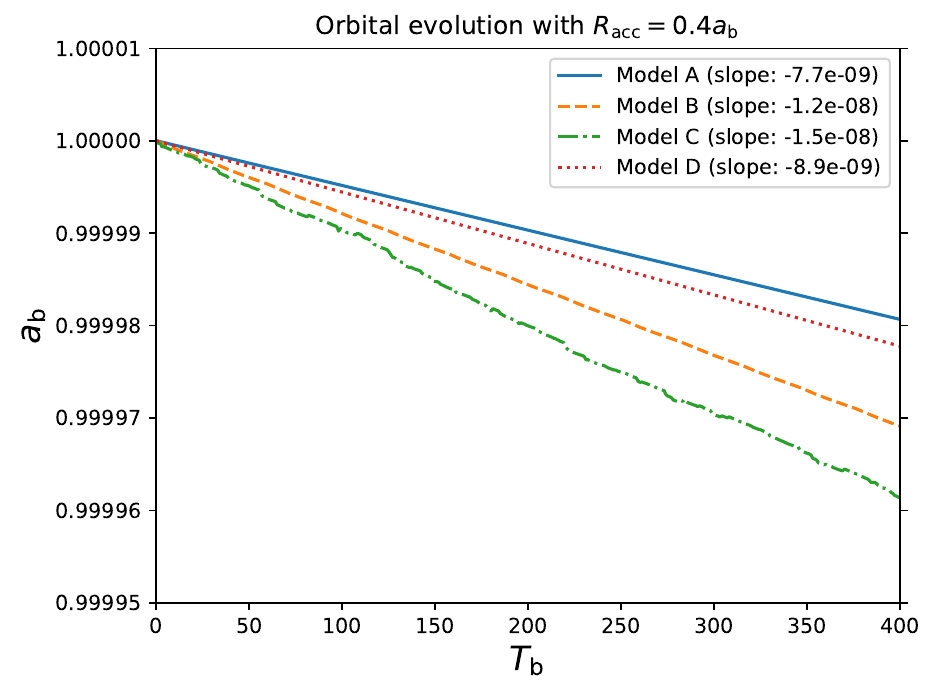}
        \includegraphics[width=8.7cm]{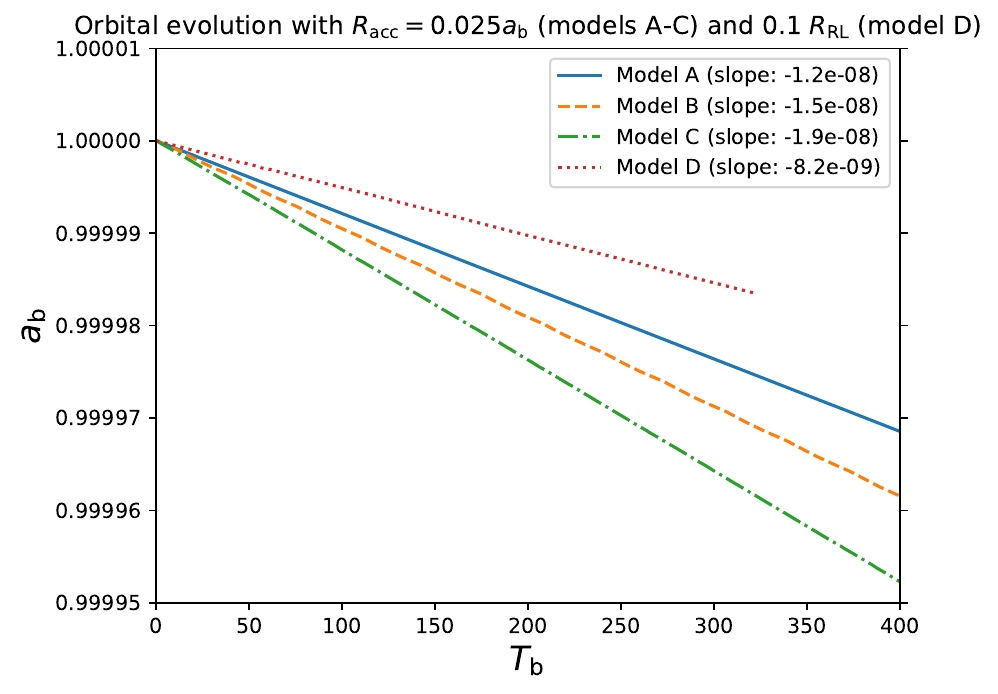}
        \includegraphics[width=8.7cm]{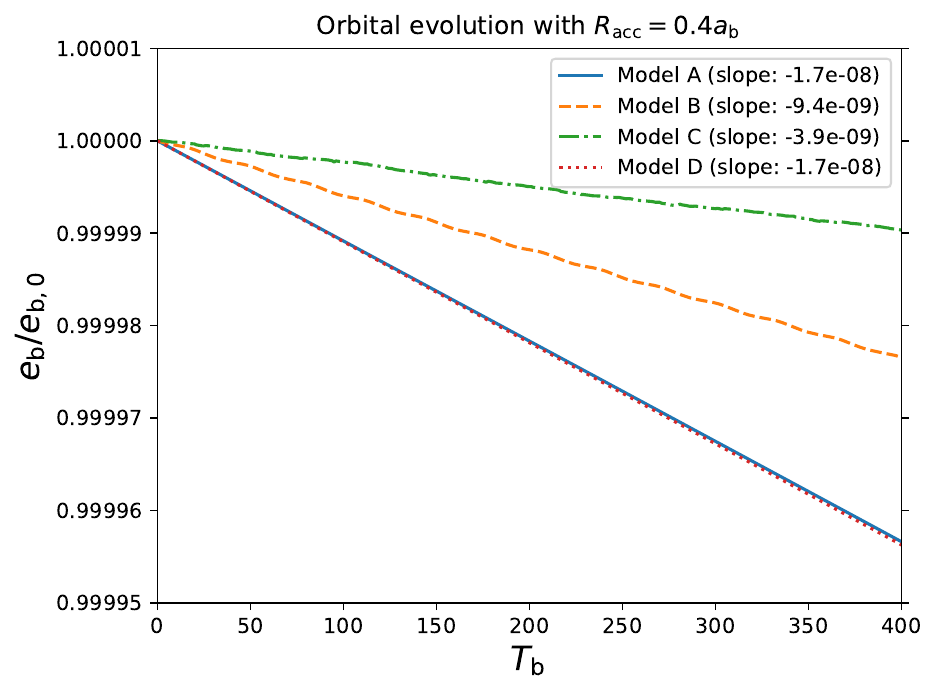}
        \includegraphics[width=8.7cm]{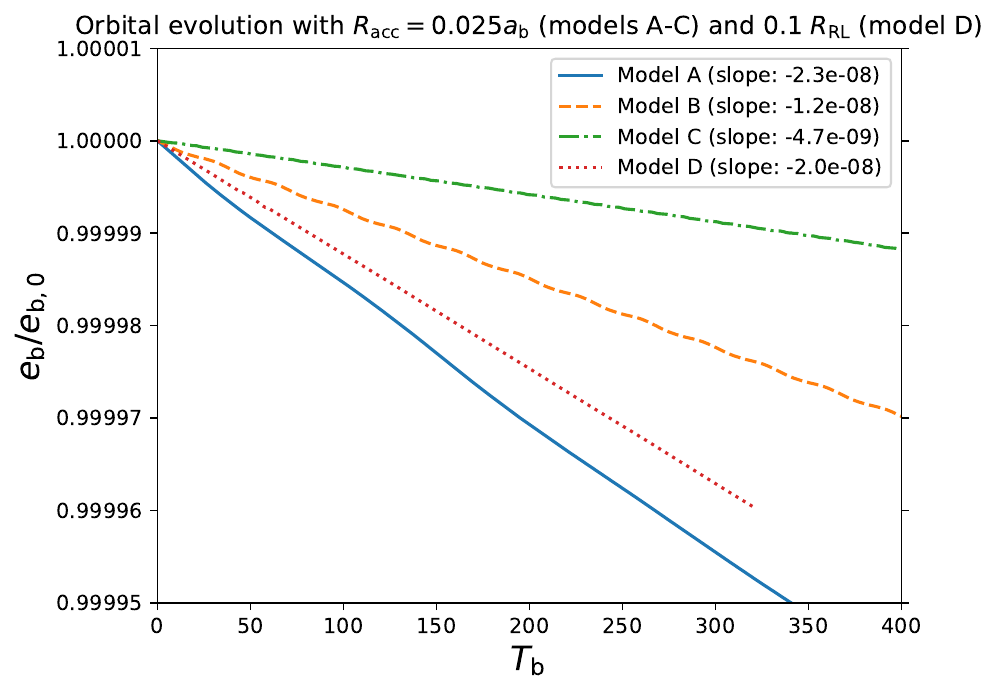}

    \caption{The semi-major axis and the eccentricity of the binary with time measured in binary orbits, $T_{\rm b}$, of models A -- D after we turn on the live binary. The blue, yellow, green and red lines represent models A, B, C and D, respectively. The left panels show cases with $R_{\rm acc} = 0.4 a_{\rm b}$ and the right panels show cases with smaller $R_{\rm acc}$. }
    \label{fig:evo}
\end{figure*}

We now turn on the back reaction so that the binary feels the gravitational influence of the disk, from which we can measure the evolution of the binary orbit. In Fig.~\ref{fig:evo} (left panels), we plot the binary semi-major axis (top left) and eccentricity (bottom left) with time, showing that all of these simulations exhibit shrinkage of the binary with the time. Model A (blue), model B (yellow dashed) and model C (green dot-dashed) exhibit shrinkage rates of $a_{\rm b}$ of -7.7$\times10^{-9}$, -1.2$\times10^{-8}$ and -1.5$\times10^{-8} \ a_{\rm b}/T_{\rm b}$, respectively and shrinkage rates of $e_{\rm b}$ of -1.7$\times10^{-8}$, -9.4$\times10^{-9}$ and -3.9$\times10^{-9} \ a_{\rm b}/T_{\rm b}$, respectively.


\subsection{Simulation results with the smaller accretion radii}
\label{sim2}
To understand the effect of $R_{\rm acc}$, we modify $R_{\rm acc}$ to 0.025 $a_{\rm b}$ and let the disks evolve until they reach steady states again (with the binary orbit fixed). In this case, the number of particles in the disk once the new steady state is reached are approximately 2.5$\times 10^6$ (model A), 2.9$\times 10^6$ (model B), and 3.5$\times 10^6$ (model C).

\begin{figure*}
    \centering
        \includegraphics[width=7.cm]{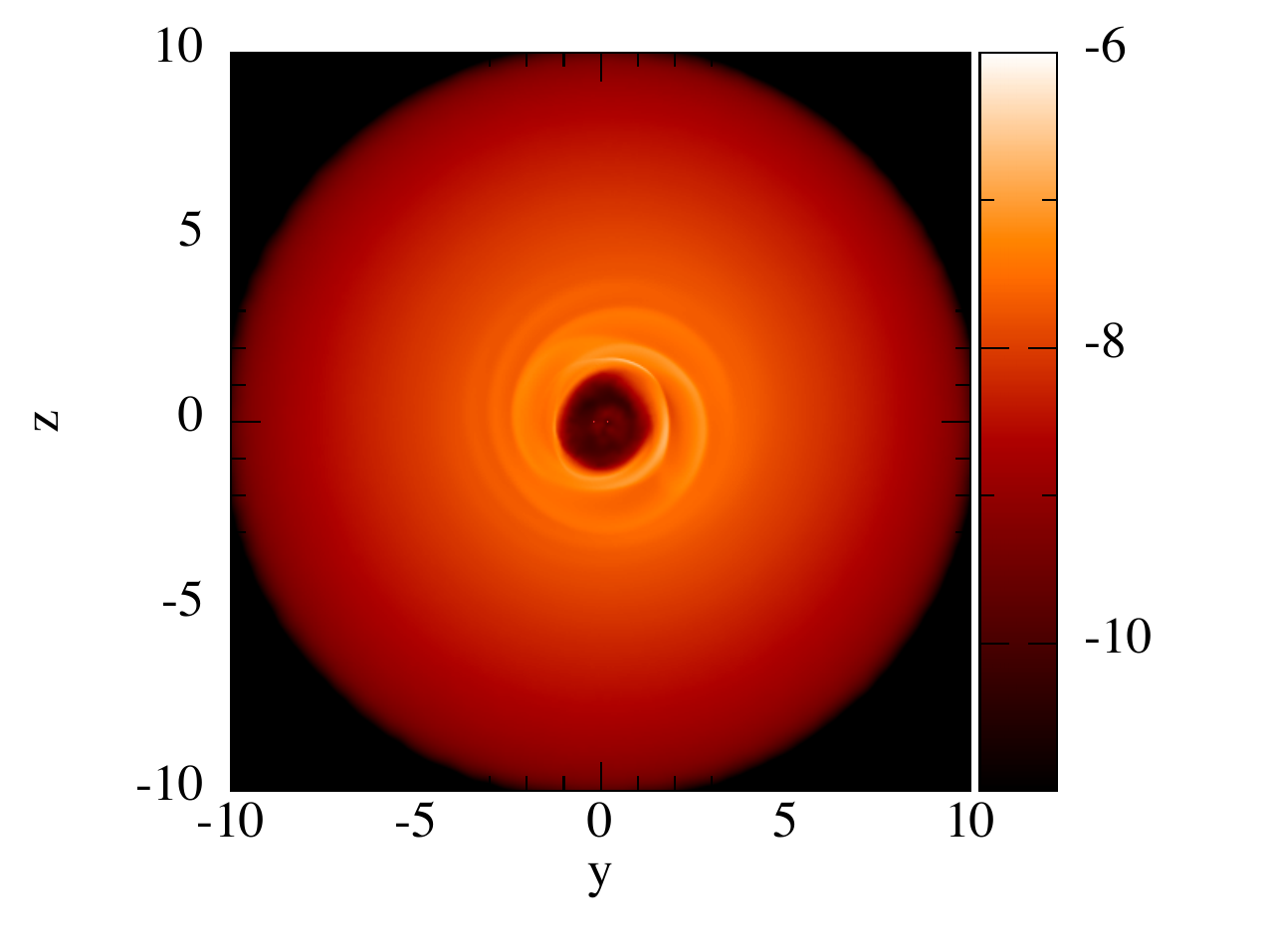}
        \includegraphics[width=7.cm]{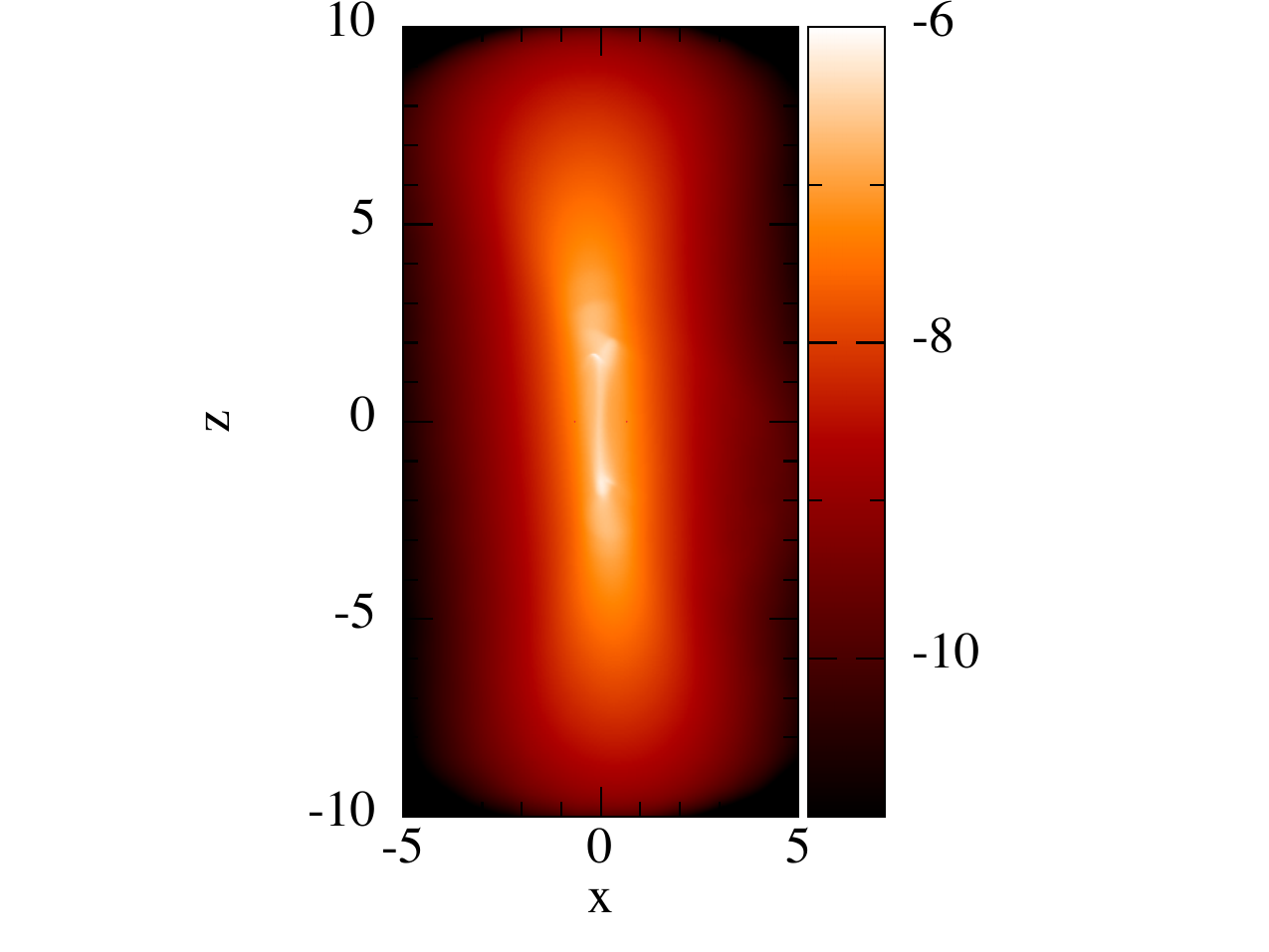}      
        \includegraphics[width=7.cm]{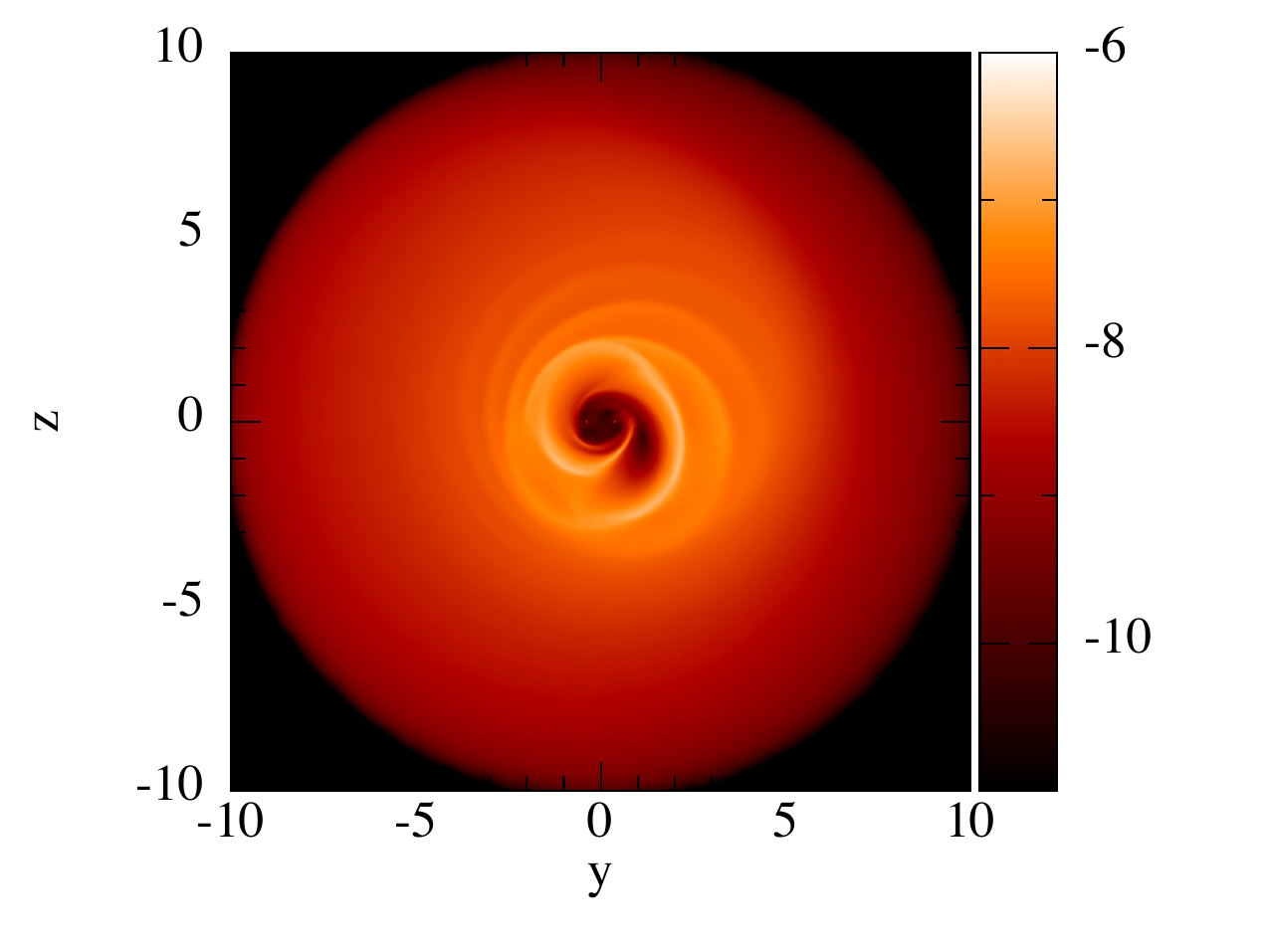}
        \includegraphics[width=7.cm]{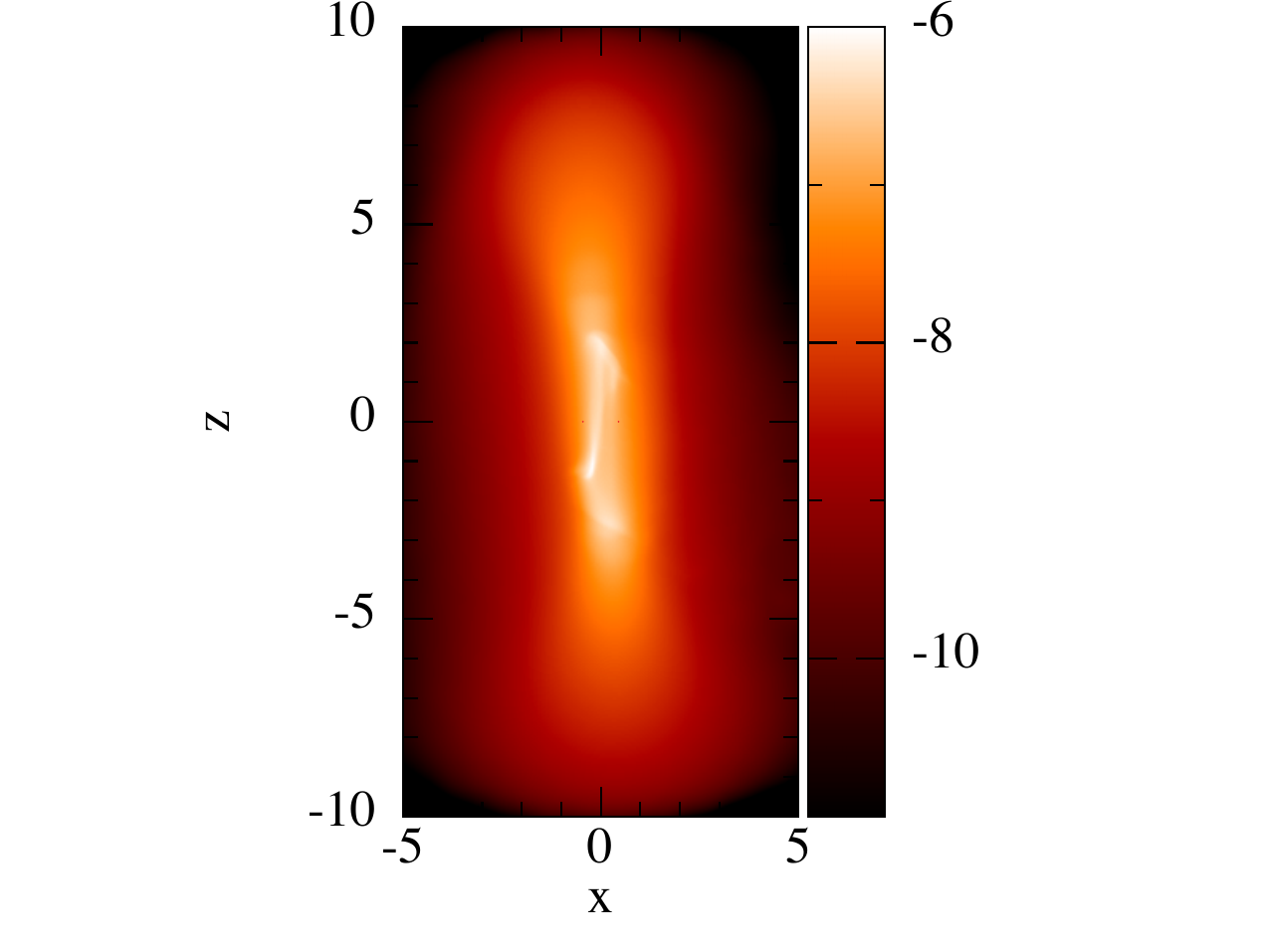}       
        \includegraphics[width=7.cm]{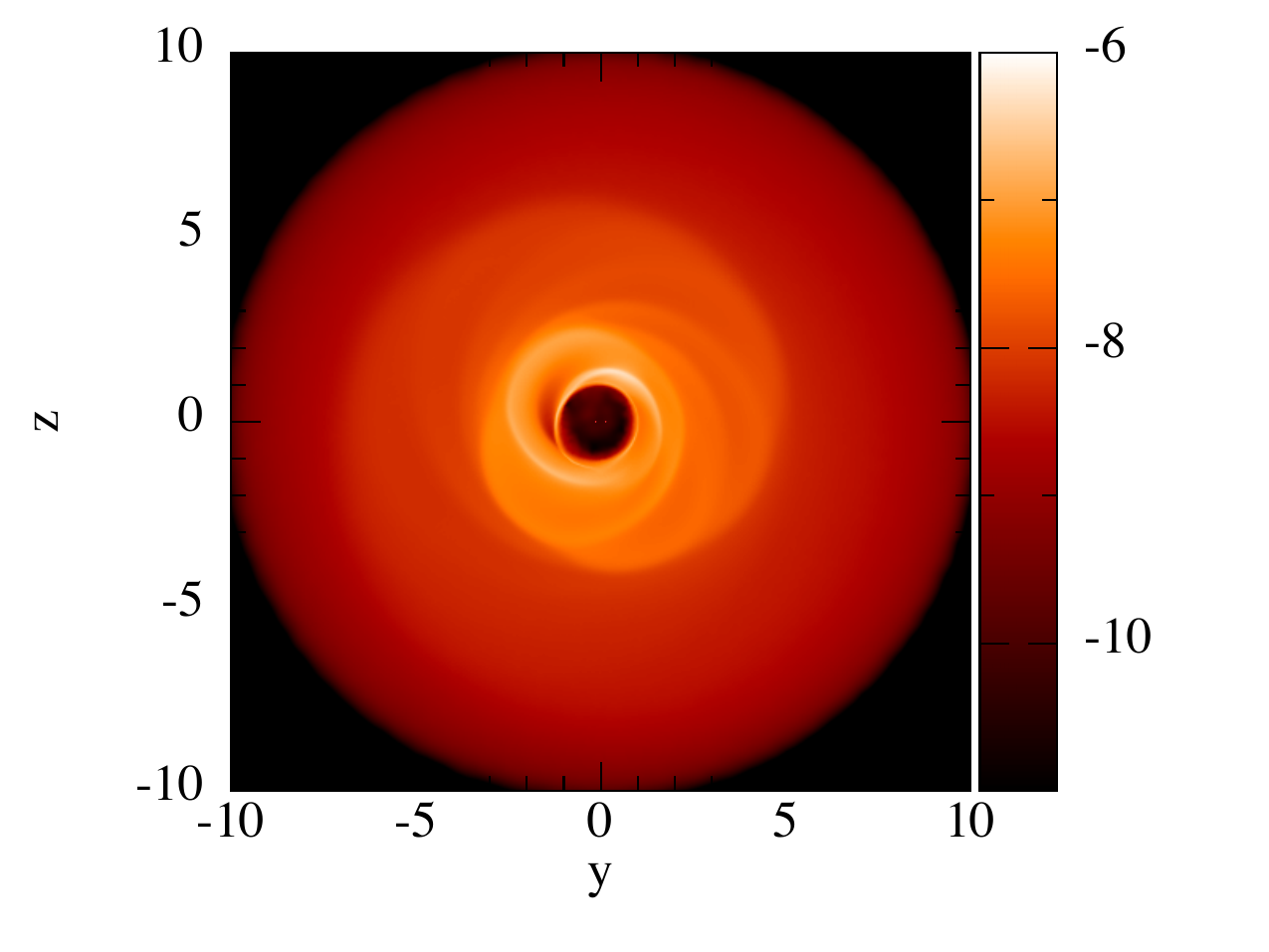}
        \includegraphics[width=7.cm]{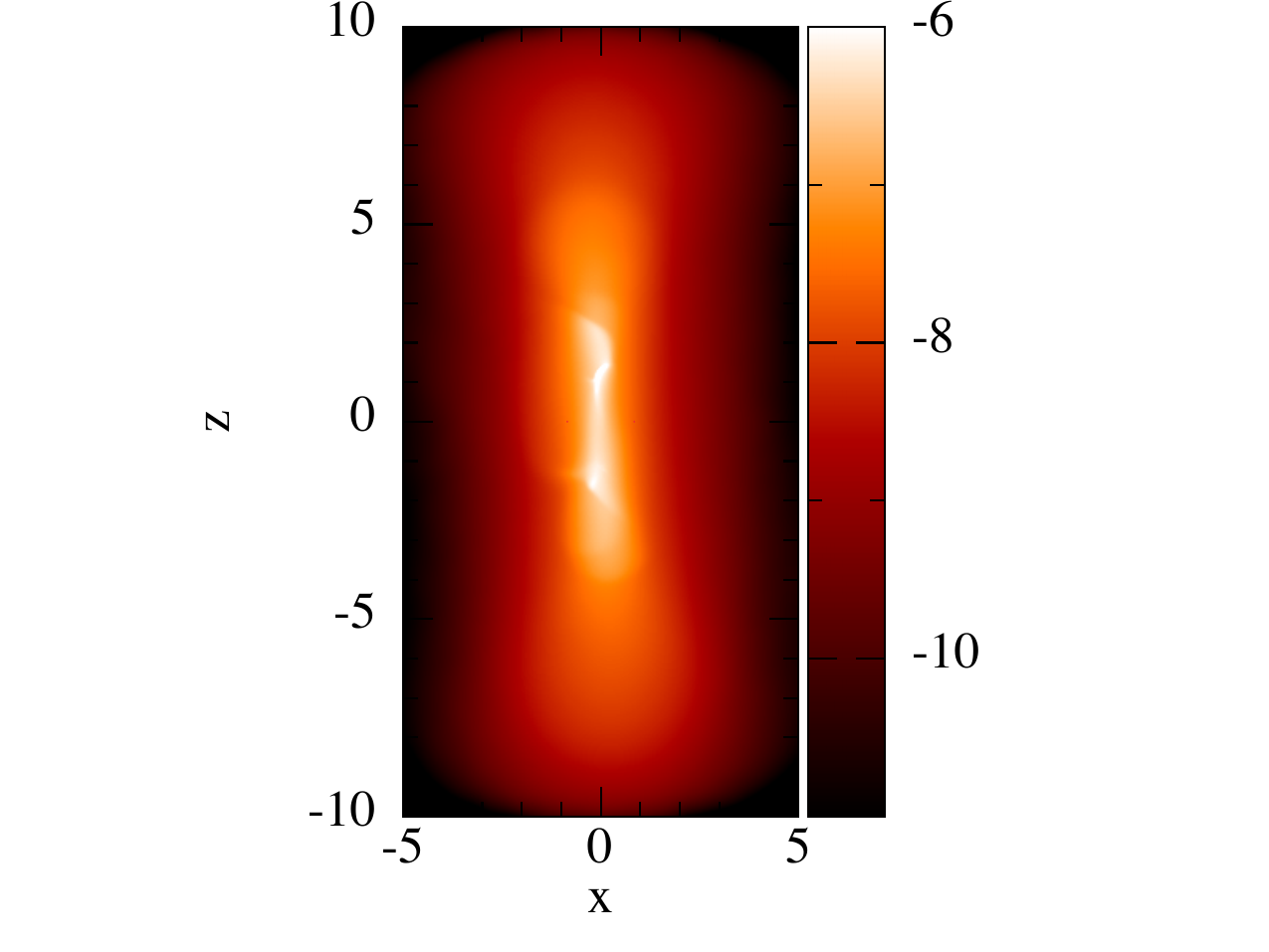}      
    \caption{Same as Fig.~\ref{fig:1} except $R_{\rm acc}=0.025\  a_{\rm b}$.}
    \label{fig:2}
\end{figure*}

In Fig.~\ref{fig:2}, we show the same column density plots as Fig.~\ref{fig:1} for the simulations with $R_{\rm acc} = 0.025 a_{\rm b}$. The two red dots representing the accretion radii of stars are significantly smaller than those in Fig.~\ref{fig:1}. However, in these simulations the mass flow rate on to the binary is sufficiently small that even with several million particles in the circumbinary disk we are not able to resolve the accretion flows around the individual stars within the binary orbit.

We show the binary orbital evolution for these cases in the right panels of Fig.~\ref{fig:evo}. Comparing with those of in the left panels, there is not much difference between them. Hence, the size of the accretion radii, $R_{\rm acc}$, does not appear to have a significant affect on the binary orbital evolution in these equal-mass binary cases.

\subsection{The effect of the binary mass ratio $q_{\rm b}$}
\label{sim3}

\begin{figure*}
    \centering
        \includegraphics[width=7cm]{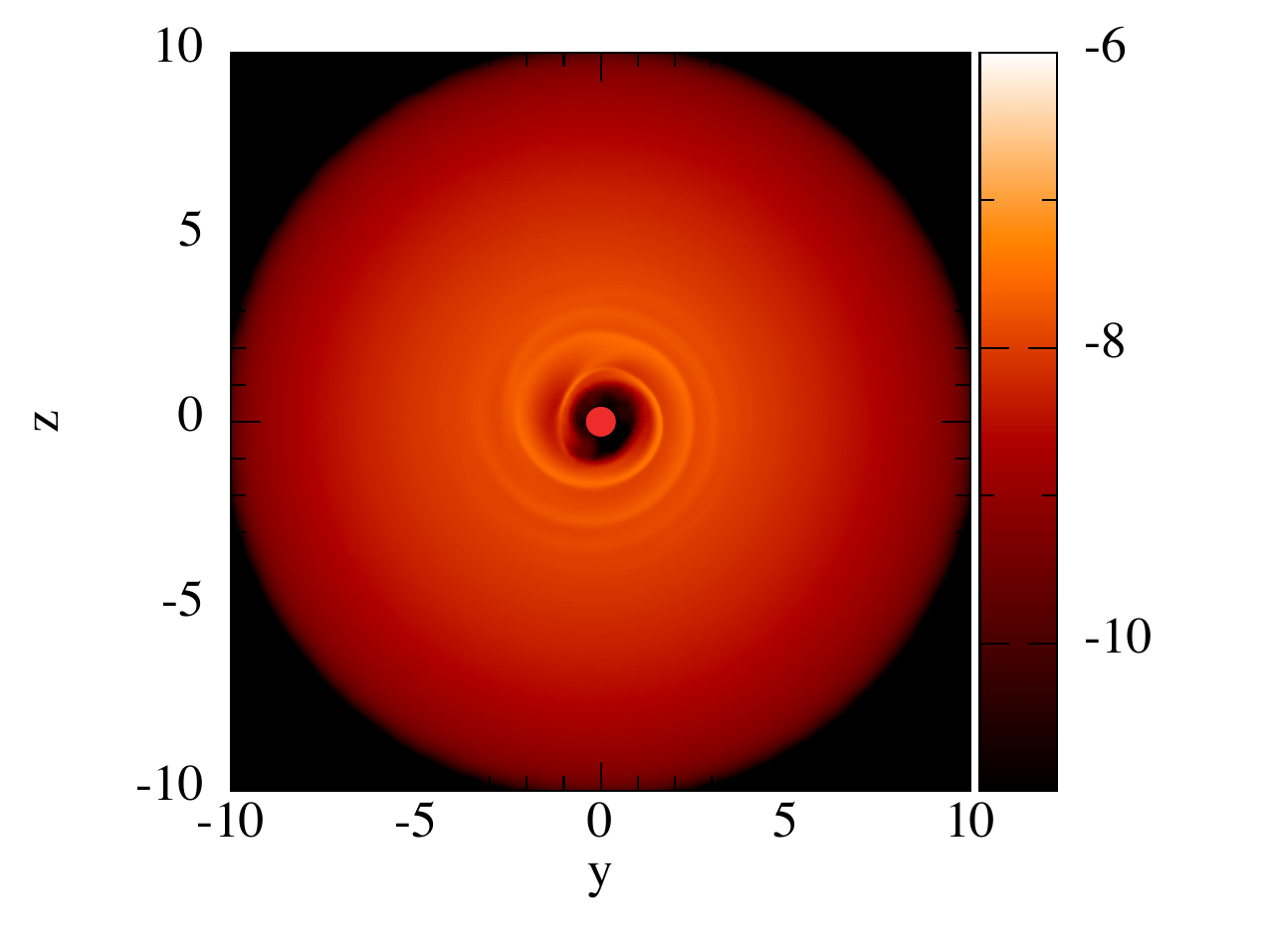}
        \includegraphics[width=7cm]{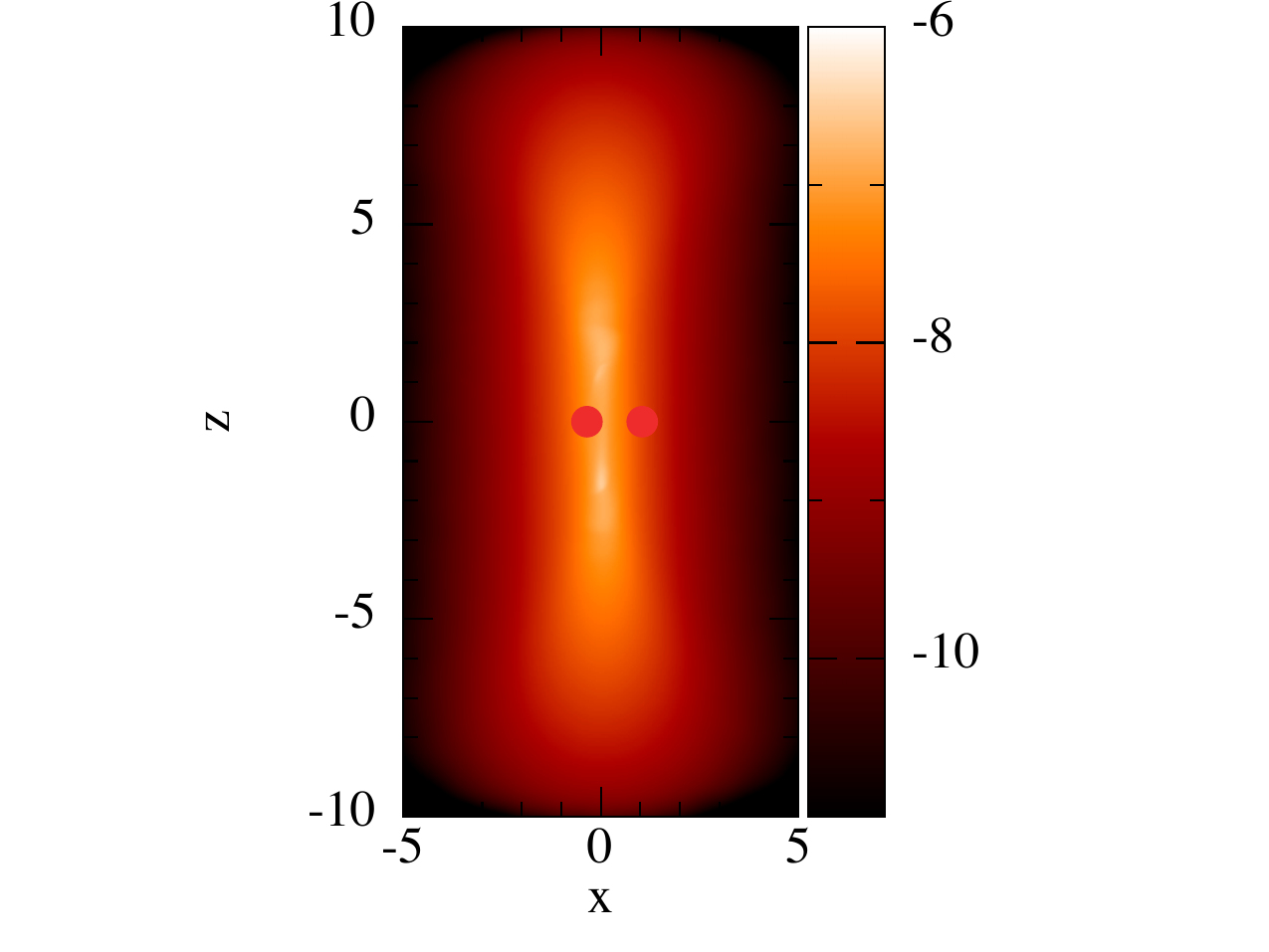}   
        \includegraphics[width=7cm]{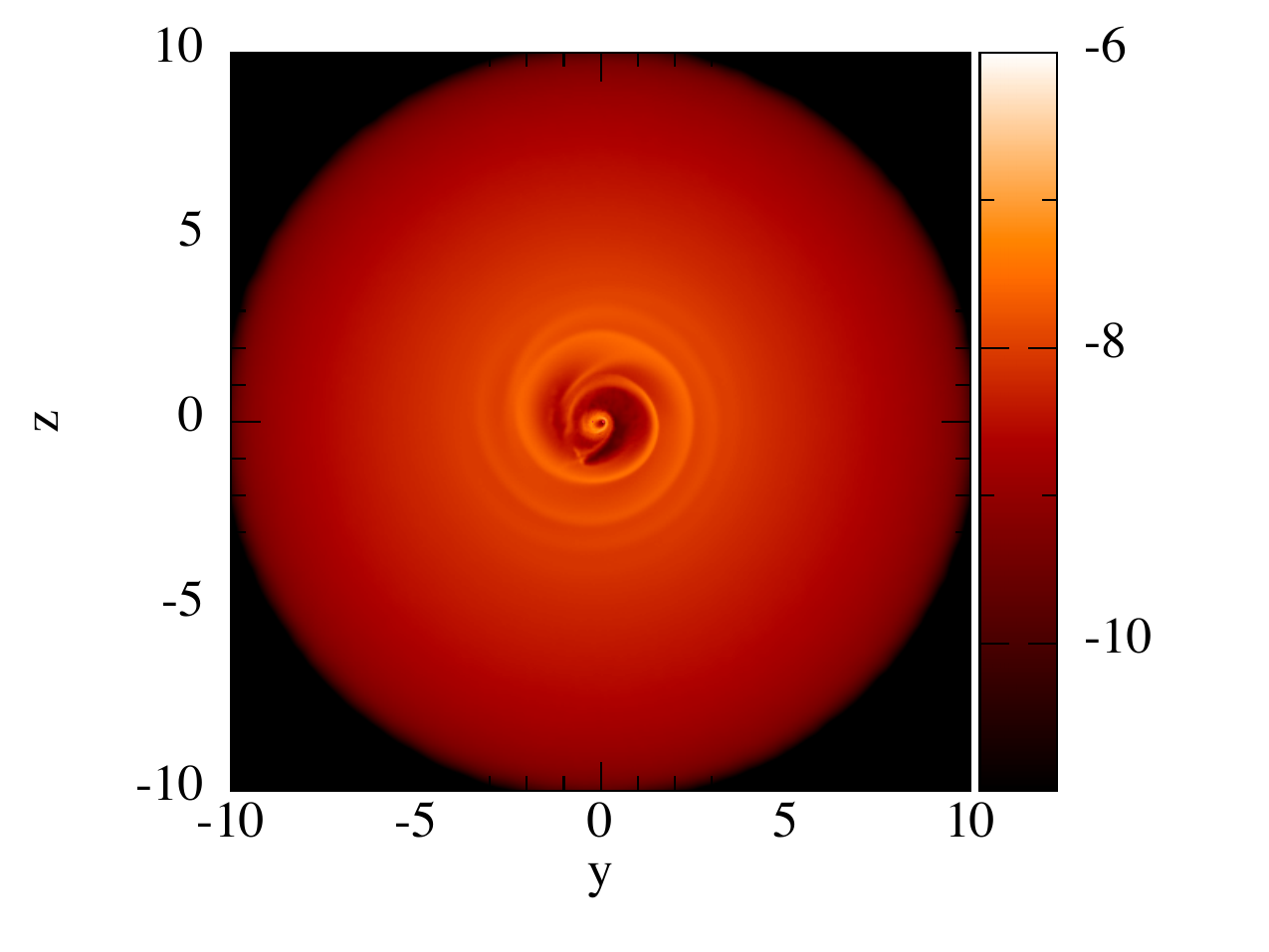}
        \includegraphics[width=7cm]{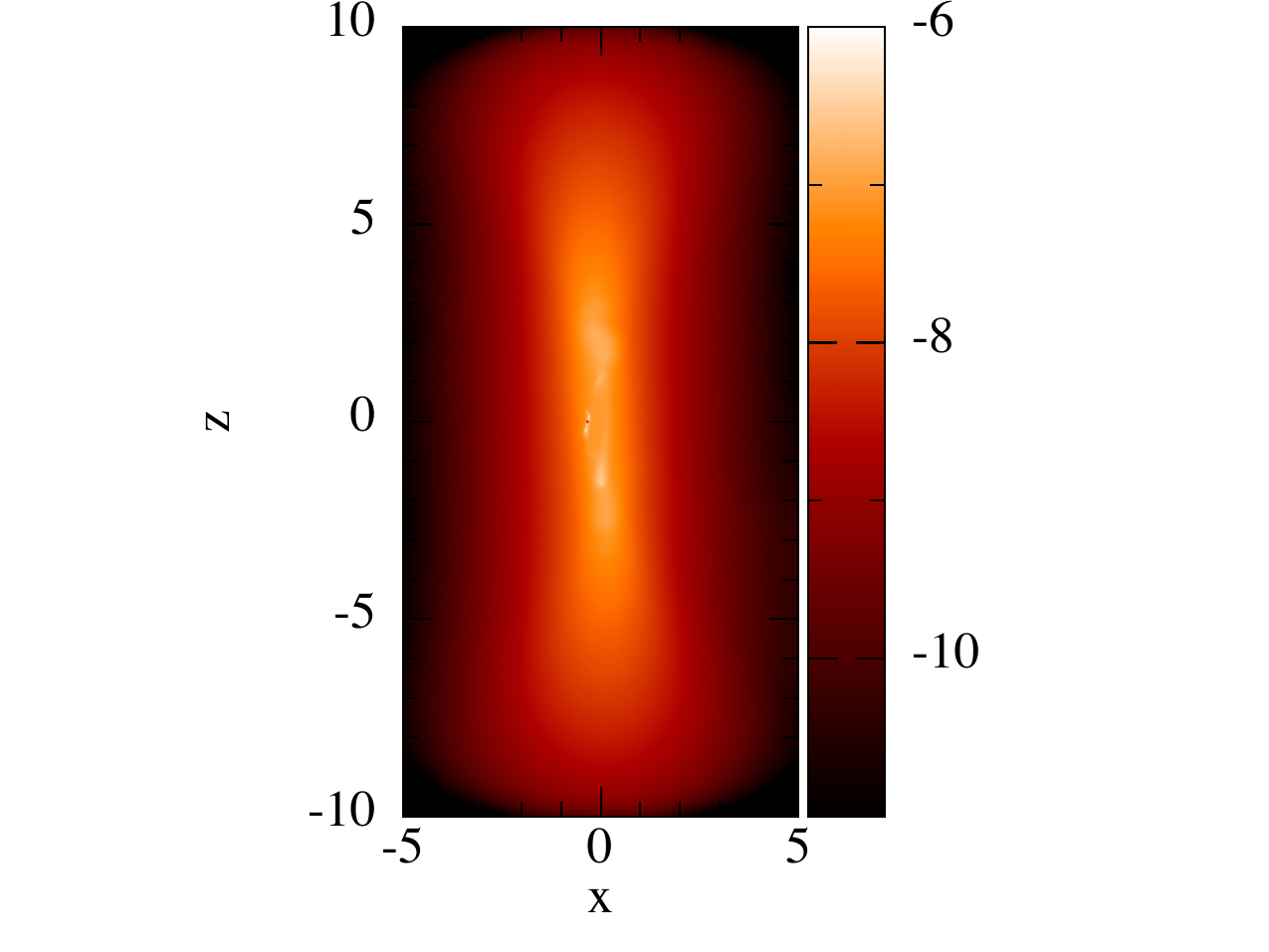}
    \caption{
    Column density plots for model D with $R_{\rm acc}$ = 0.4 $a_{\rm b}$ (top panels) and $0.1\times R_{\rm RL}$ (lower panel) as they reach the steady state. Two red circles represent the location of the binary stars (sink particles) with sizes equal to the size of the accretion radius of each star. The unit of the axis is $a_{\rm b}$ and the color bars are the same for all of the panels with the density in arbitrary units.}
    \label{fig:3}
\end{figure*}

In this subsection, we consider a disk with the same parameters as above with $e_{\rm b} = 0.4$, except this time with a lower $q_{\rm b} = 1/3$. In our first simulation (top panels of Fig.~\ref{fig:3}), we assume both the primary and secondary have $R_{\rm acc}$ = 0.4 $a_{\rm b}$. As this model reaches a steady state, it has approximately $1.6\times10^6$ particles in the disk. As with the higher mass ratio case, the binary excites density waves in the disk, and the red dotted line in the left panel of Fig.~\ref{fig:evo} shows that the binary's separation shrinks over time with a rate that is similar to that of model A.

As before, we again reduce the accretion radii and relax the disk to the corresponding new steady state. However, for this case of an unequal mass ratio we set the accretion radii as a fraction (0.1) of the Roche lobe radius for each star (calculated as if the binary were circular; i.e. using the binary semi-major axis). Specifically, we set the accretion radii to $0.1 R_{\rm RL}$, where $R_{\rm RL}$ is defined by \citep{Eggleton1983}
\begin{equation}
    R_{\rm RL} = \frac{0.49 q_{\rm b}^{2/3}}{0.69 q_{\rm b}^{2/3}+\ln({1+q_{\rm b}^{1/3}})} a_{\rm b}.   
\end{equation}
Thus, $R_{\rm acc}$ of the primary and the secondary are 0.0476 $a_{\rm b}$ and = 0.0289 $a_{\rm b}$, respectively. 

After reducing $R_{\rm acc}$ for both stars, and allowing the system to reach a steady state again, it still has $N_{\rm p} \approx 1.6\times 10^6$ particles. In the lower panels of Fig.~\ref{fig:3}, a circumstellar disk forms around the primary star with approximately 7000 particles (equivalent to a mass of $\approx 7\times10^{-9}M_{\rm b}$) within the Roche lobe and the accretion rate onto the primary is about 3.5$\times 10^{-10}M_{\rm b}/T_{\rm b}$. It is possible for the disk to form in this case due to the larger Roche lobe size, which can support more long-lived material.

\begin{figure}
    \centering
        \includegraphics[width=8.5cm]{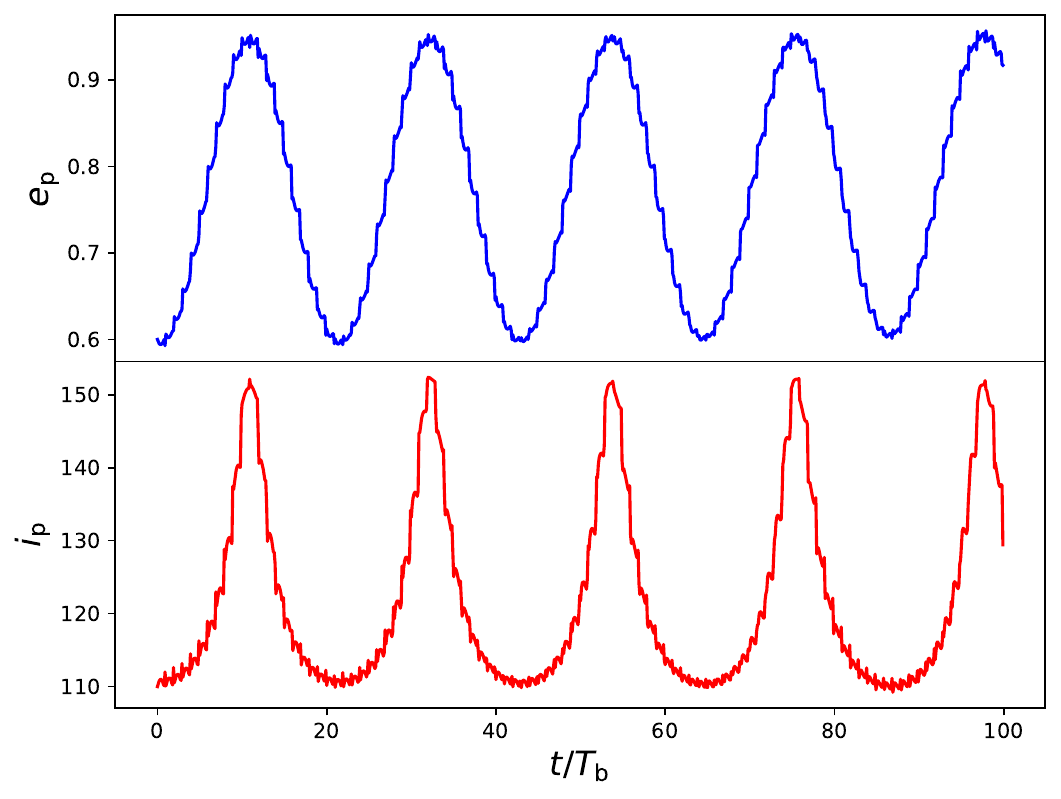}
    \caption{The time evolution of the test particle's $e_{\rm p}$ (top panel) and $i_{\rm p}$ (lower panel) around the CPD with the same binary parameters to model D. }
    \label{fig:testp}
\end{figure}

\begin{figure}
    \centering
        \includegraphics[width=8.5cm]{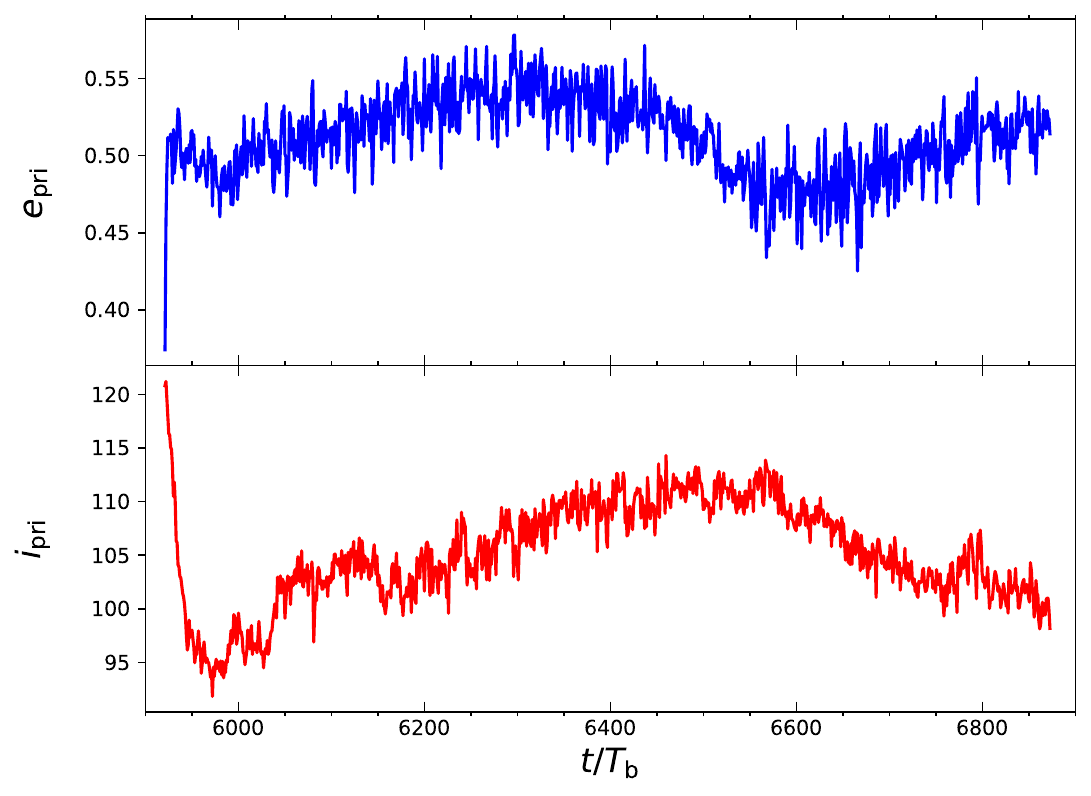}
    \caption{Time evolution plots for $e_{\rm pri}$ (top) and $i_{\rm pri}$ of the circumprimary disk formed in model D with the smaller $R_{\rm acc}$. This plot starts as we reduce the accretion radii and for the first few orbits of the binary the circumprimary disc is being formed. After this it settles into a configuration with $e_{\rm pri} \approx 0.5-0.55$ and $i_{\rm pri} \approx 100-110^\circ$. While there is some variability, the ZKL cycles in Fig.~\ref{fig:testp} are not clearly present.}
    \label{fig:cpd}
\end{figure}

The disk around the primary seems stationary in our simulations. To check whether this is expected, and whether this is the case, we plot in Figure~\ref{fig:testp} and Figure~\ref{fig:cpd} the eccentricity (top panel) and inclination (bottom panel) values of a test particle orbiting in the binary potential at the location of the disk (Fig.~\ref{fig:testp}) and the equivalent plots for the disk in the simulation (averaged over the particles that make up the disk; Fig.~\ref{fig:cpd}). For the test particle case, we use the $N$-body code, {\sc rebound} with the {\sc IAS15} integrator \citep{Rein2015a} to model a test particle on an orbit with initial parameters $e_{\rm p} = 0.6$ and $i_{\rm i}=110^{\circ}$ at 0.1$a_{\rm b}$ from the primary. The test particle in Figure~\ref{fig:testp} clearly shows von Zeipel-Kozai-Lidov oscillations \citep[ZKL,][]{vonZeipel1910,Kozai1962,Lidov1962} with a period of about 23 $T_{\rm b}$. However, while the circumprimary disk in the simulation does exhibit some variations in inclination and eccentricity, these are neither at the expected ZKL frequency nor are they anti-correlated.

Fig.~\ref{fig:cpd} shows that the circumprimary disk is long-lived, lasting for the duration of the simulation ($\approx 1000 T_{\rm b}$ while $e_{\rm pri}$ ranges from 0.45 to 0.55 and $i_{\rm pri}$ ranges from $100^{\circ}-110^{\circ}$. With higher resolution, and a lower viscosity, we would expect ZKL oscillations to be present. However, here the accretion timescale ($\approx 20 T_{\rm b}$) is shorter than the ZKL timescale, and thus particles pass through the disk without undergoing the (coherent) cycles observed in, e.g., \cite{Dogan2015,Martinetal2014b}.

Additionally, we note that with a resolved circumstellar disk around the primary, the binary's orbit still shrinks with the time (see the right panel of Fig.~\ref{fig:evo}). The rate at which the binary shrinks is marginally lower than that of with $R_{\rm acc}=0.4$ by around 10\%.

\section{Discussion}
\label{diss}

\subsection{The mass and the resolution of the circumbinary disk}
In our simulations, we consider relatively small mass of the disk, because the polar stationary inclination $i_{\rm s}$ depends on $e_{\rm b}$ and the ratio of the angular momentum of the disk to the angular momentum of the binary \citep[see equation 17 in][see also \citealt{Aly2015}]{MartinandLubow2019}. Hence, a massive disk or a massive particle result in $i_{\rm s} < 90^{\circ}$ \citep[e.g.][]{Chen20192, Abod2022, Smallwood2023}. Since the direction of the mass injection function in our models is perpendicular to the initial disk plane, we do not want the deviation of the injection angle during the disk evolution. Hence, the smaller disk mass can prevent further evolution of $i_{\rm s}$ with the smaller mass injection rate.

\begin{figure}
    \centering
        \includegraphics[width=8.3cm]{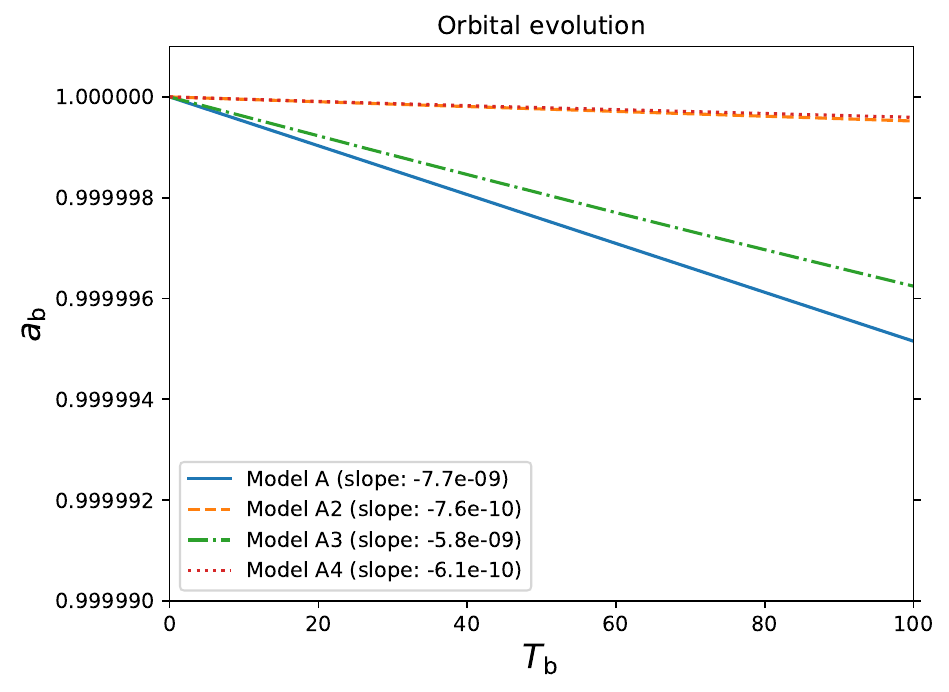}
    \caption{The separation of the binary with $T_{\rm b}$ of models A (blue), A2 (orange, lower $M_{\rm d}=10^{-7}M_{\rm b}$), A3 (green, lower $N_{\rm p, ini}=10^5$) and A4 (red, lower $N_{\rm p, ini}=10^5$ and lower $M_{\rm d}=10^{-7}M_{\rm b}$). The lines for models A2 and A4 nearly overlap.}
    \label{fig:A23}
\end{figure}

Additionally, the net effect of the binary's orbital evolution is scaled to the mass of the disk in general. Thus, we consider model A2, which has same setup to model A except lower disk mass ($M_{\rm d}=10^{-7} M_{\rm d}$). We also run additional two models A3 and A4 which have the same setup as model A except lower resolution ($N_{\rm p, ini}=10^5$) for model A3 and lower resolution  ($N_{\rm p, ini}=10^5$) and mass ($M_{\rm d}=10^{-7} M_{\rm d}$) for model A4. After models A3 and A4 reach steady states, their $N_{\rm p}$ are about $1.5 \times 10^5$. In Fig~\ref{fig:A23}, we show that the time evolution of $a_{\rm b}$ including models A (blue line), A2 (yellow dashed line), A3 (green dot-dashed line) and A4 (red dotted line). 

Models A and A2 show that the orbital evolution rate is scaled to the disk mass, $M_{\rm d}$, in our simulations. On the other hand, comparing with models A and A3, the effect of the resolution can affect the shrinkage rate by about 20 per cent, with the faster shrinkage rate occurring at higher resolution. Furthermore, the shrinkage rates of models A3 and A4  also display the scaling with $M_{\rm d}$. Hence, if we ignore the other effects, for the more massive polar disk ($M_{\rm d}=10^{-3}$ or $10^{-4} M_{\rm b}$), the net shrinkage rate for each model would be significantly higher than the rates shown in our figures.

\subsection{Orbital evolution with prograde and retrograde disks}
A binary accretes different angular momentum components depending on the specific alignment of the circumbinary disk, such as prograde, polar, or retrograde. The different specific angular momentum deposited onto the binary may result in diverse orbital evolution. In this subsection, we consider Model A with the similar setup and the prograde CBD and the retrograde CBD to do the comparison. As prograde and retrograde CBDs reach steady states, their $N_{\rm p}$ increase from 10$^6$ to 1.52$\times$ 10$^6$ and 1.27$\times$ 10$^6$, respectively. Their $m_{\rm d}$ also increase with increasing $N_{\rm p}$. 

\begin{figure}
    \centering
        \includegraphics[width=8.3cm]{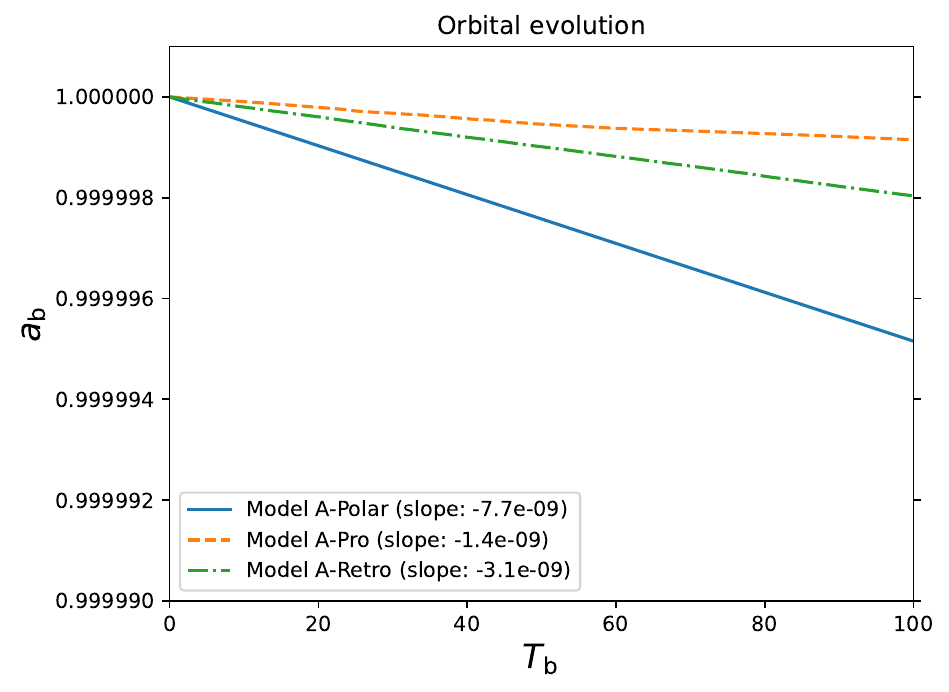}
    \caption{The semi-major axis of the binary with time measured in binary orbits, $T_{\rm b}$, of model A after we turn on the live binary with $N_{\rm p, ini}=10^6$ including polar (blue), prograde (yellow dashed) and retrograde (green dash-dotted) CBDs.}
    \label{fig:A_prore}
\end{figure}

In Fig.~\ref{fig:A_prore}, we show the time evolution of model A with the polar (blue), the prograde (yellow dashed) and the retrograde (green dash-dotted) CBDs. The binary with the prograde CBD has the slowest shrinkage rate because the binary acquires the specific angular momentum in positive z-component from the prograde CBD. On the other hand, the binary with the retrograde CBD has the middle shrinkage rate. In this case, although the binary acquires the specific angular momentum in negative z-component, the Lindblad resonances are significantly weakened in this case \citep[see][for the details]{Nixon2015}. Hence, the binary's orbital angular momentum cannot be transferred to the CBD efficiently for the parameters simulated here. 

Finally, simulations in Fig.~\ref{fig:1} and Fig.~\ref{fig:3} show that Lindblad resonances still occur in polar CBDs. Moreover, the binary acquires less specific angular momentum in positive z-component from the polar CBD than from the prograde CBD. As a result, the binary with the polar CBD has fastest shrinkage rate. The exact torques acting in the system, and thus which of prograde, retrograde or polar provides the fastest shrinkage rate, depends on the simulated parameters. 

\subsection{ZKL oscillations of the circumprimary disk}
\label{dis:ZKL}
In model D, a mini disk around the primary forms after reducing $R_{\rm acc}$ to 0.1 $R_{\rm RL} \simeq 0.0476 a_{\rm b}$ . Since particles accreted onto primary have near--polar inclinations, inherited from their circumbinary orbits, the circumprimary disk (CPD) is both highly inclined and highly eccentric.\footnote{Typically, disks  that form within a binary are only eccentric towards their outer edges and are efficiently circularized. That this does not happen here, we attribute to the continued supply of eccentric material and the absence of strong apsidal precession of the disk orbits which is the main route to circularization in the planar case. See Figs. 6 \& 7 of \cite{Nixonetal2013} for examples.} In Fig~\ref{fig:cpd}, the mini disk maintains $e_{\rm pri}$ about 0.45 -- 0.55 and $i_{\rm pri}$ near polar ($95^{\circ}$--$110^{\circ}$) for $>1000 T_{\rm b}$.

In previous SPH simulations of CBDs presented by \citet{Smallwood2023}, a CPD is found to form that does undergo a small number of ZKL oscillations before settling into a non-oscillating state after around $50$ binary orbits, but the damping was much weaker at the higher resolution (see their Fig 6), and hence they concluded that these oscillations can be long-lived.\footnote{ZKL oscillations were also present in some of the CBD simulations of \cite[][see their Figs. 6 \& 7]{Nixonetal2013}.} Their simulation is able to produce a CPD at lower resolution than ours due to the application of a reduced temperature in the disk regions around the star.\footnote{We thank Steve Lubow for drawing this to our attention.} A reduced temperature, and thus sound speed, might be expected to restrict coherent ZKL oscillations, which requires efficient communication of the precession across the disk \citep{Martinetal2014b}, but since the CPD formed is radially narrow (see Fig.~11 of \citealt{Smallwood2023}) and has smoothing lengths of order the disk thickness, the oscillations were communicated efficiently. We suggest that the decay of the ZKL oscillations in their simulations arises, in part, due to the slowing of the mass transferred from the CBD with time, allowing the disk to align/circularize somewhat over the course of the simulation. It is also worth comparing Figures 6 \& 8 in \citet{Smallwood2023} which show that the different ZKL oscillations around CBDs at varying levels of misalignment. We therefore speculate that higher-resolution simulations that can create a well-resolved CPD from a near-polar CBD, could exhibit long-lived ZKL oscillations, particularly if the physical viscosity employed were significantly lower than we employ here.

Alternatively, as mini disks in a real situation should have a similar (or greater) sound speed to the CBD which supplies accretion materials from streams, it might be that the accretion timescale through the mini disks is small enough to prevent the formation of coherent ZKL cycles. In our simulation, ZKL oscillations are not obvious as $H/R=0.1$. In this case, the near-polar mini disks may provide favorable conditions for the in-situ formation of highly misaligned S-type planets. Misaligned planets orbiting one star in binary systems are relatively common. The Transiting Exoplanet Survey Satellite (TESS) Objects of Interest project has confirmed 56 out of 170 binary systems as hosting planets. Among these, 73$\%$ exhibit Kepler-like architectures ($R_{\rm p} \leq R_{\oplus}$, $a < 1$ au) with a typical mutual inclination of $35^\circ \pm 24^\circ$ between the planet and its host star. In contrast, 65$\%$ of systems hosting close-in giant planets ($P < 10$ days, $R_{\rm p} > R_{\oplus}$) have typical mutual inclinations of $89^\circ \pm 21^\circ$ \citep{Behmard2022}.

\subsection{Polar planet formation}

Observations of solar-type binaries reveal a distinct break in the period–eccentricity relation for 127 binaries at an orbital period of 12 days \citep[see Figure 14 in][]{Raghavan2010}. Binaries with orbital periods of $\lesssim 12,\rm days$ are nearly circular, whereas those with periods $\gtrsim 30,\rm days$ exhibit diverse eccentricities. 

This result is consistent with N-body simulations of a binary with a polar debris CBD presented in \citet{Chen2024b}, which show that for sufficiently small binary separations ($\leq 0.2$ au), tidal interactions can efficiently shrink the orbit leading to the merger of the binary star system. Hence, our SPH simulations in this study provide a pathway for reducing the binary separation before tidal interactions dominate the system's orbital evolution.

On the other hand, the rate of the binary's orbital evolution is scaled to the disk mass, and protoplanetary disks can reach masses of $10^{-2} \sim 10^{-3} M_{\odot}$ and persist for up to 10 Myr for low mass stars \citep{Williams2011}. As a result of the binary-polar CBD interaction, a binary with a large initial separation could merge within a disk lifetime, leaving a polar disk around a single star and polar planets can form inside the polar disk, as suggested by \citet{Chen2024b}. 

\subsection{Supermassive black hole binaries with polar disks}
Recent observations have identified many candidate supermassive black hole (SMBH) binaries \citep[e.g.][]{Fabbiano2011, Kharb2017, Balmaverde2018, Bhatta2018, Benitez2019, Foord2019, Kharb2019, Sebastian2019, Wang2020, Li2021, Nandi2021, Cheng2024,  Mondal2024, Rigamonti2025}. The mechanisms that cause SMBH binaries to merge remain unclear: interactions with stars and gas in the galaxy can bring the SMBH to separations of order a parsec, while gravitational waves are only efficient at merging the binary on smaller scales of $10^{-4}-10^{-3}$\,pc. The existence of gap between these scales is known as the final parsec problem \citep{Begelmanetal1980,MM2001}. Gas disks provide one  mechanism to overcome this gap \cite[e.g.][]{Armitage2002b,Mayeretal2007, Dottietal2007, Cuadra2009} and retrograde \citep{Nixonetal2011a,Nixon2015} or misaligned disks  \citep{Nixonetal2013} may be promising routes to shrinking the binary on cosmologically short timecales. However, these routes naturally lead to large binary eccentricities \citep[e.g.][]{Roedig2011}. In an environment where there is chaotic gas infall to galaxy centers \citep{king:2006,king:2007,Nixonetal2011b} polar CBDs should therefore be common in SMBH binaries \citep[e.g.][]{Aly2015,Childs:2024,Martin:2024}. On the other hand, general relativity (GR) can suppress polar alignment if the GR-induced precession timescale is much shorter than the binary-driven secular precession timescale depending on the disk parameters \citep{Childs:2024}. Here we have shown that interactions with a polar CBD lead to shrinkage of the binary orbit over time. We therefore expect that polar CBDs may play an important role in the evolution of SMBH binary orbits and should be considered as part of evolutionary models for SMBH binaries \citep[e.g.][]{Siwek:2024}.

\section{Conclusion}
\label{con}
We have investigated the interaction between an eccentric binary and a polar circumbinary disk using SPH simulations. The back reaction and the mass accumulation from accretion onto the binary are neglected until the disk has achieved into a steady state structure. This approach allows an accurate determination of the binary evolution, removing dependence on the initial disk conditions or the time at which the properties are measured. Our simulations show that the net effect of binary-disk interaction in the polar case is the shrinkage of the binary orbit over time. We find the eccentricity of the binary has only a minor effect on the shrinkage rate. Moreover, we consider different sizes for the accretion radius for both stars of the binary, but again the shrinkage rate does not vary significantly in all cases. For unequal mass binaries we found that a mini disk can form in the simulation and sustain its eccentricity and inclination for a long timescale (as long as the simulation was run for; $ > 1000 T_{\rm b}$) without ZKL oscillations. We speculate that this might form conditions conducive to the formation of S-type planets in polar orbits. We also suggest that polar CBDs may play an important role in the merger of SMBH binaries in galaxy centers.

\begin{acknowledgements}
We thank Steve Lubow for useful discussion about ZKL oscillations and John Ruan for helpful comments on an initial draft. CC and CJN acknowledge support from the Science and Technology Facilities Council (grant number ST/Y000544/1). CC acknowledges support from the CITA National Fellowship (grant number DIS-2022-568580). CJN acknowledges support from the Leverhulme Trust (grant number RPG-2021-380). This project has received funding from the European Union’s Horizon 2020 research and innovation program under the Marie Skłodowska-Curie grant agreement No 823823 (Dustbusters RISE project). CC gratefully acknowledges the hospitality of Stony Brook University and the Flatiron Institute during a Dustbusters secondment. We thank Xiao Hu for sharing HPC time for us to run simulations by an XSEDE allocation (AST200032). This work was performed using Anvil \citep{McCartney2014}, operated by the Purdue University (https://er.educause.edu/articles/2014/7/empowering-faculty-a-campus-cyberinfrastructure-strategy-for-research-communities).
\end{acknowledgements}

\bibliography{main}
\bibliographystyle{aasjournal}

\end{document}